\documentclass[preprint2]{aastex}

\usepackage{graphicx}

\newcommand{\ec}{$\eta$~Car}
\newcommand{\lm}{$\lambda$}

\newcommand{\degree}{\ensuremath{^\circ}}

\begin{document}

\title{High-excitation emission lines near eta Carinae, and its likely companion star\altaffilmark{1}}

\author{Andrea Mehner\altaffilmark{2}, Kris Davidson\altaffilmark{2}, Gary J.~Ferland\altaffilmark{3}, Roberta M.~Humphreys\altaffilmark{2}}
 
\altaffiltext{1}{This research was supported by grants no. GO-9973, 10844, and 11291 from the Space Telescope Science Institute. The HST is operated by the Association of Universities for Research in Astronomy, Inc., under NASA contract NAS5-26555.} 

\altaffiltext{2}{Department of Astronomy, University of Minnesota, Minneapolis, MN 55455}

\altaffiltext{3}{Department of Physics and Astronomy, University of Kentucky, Lexington, KY 40506}

\email{mehner@astro.umn.edu, kd@astro.umn.edu, gjferland@googlemail.com, roberta@aps.umn.edu}

\begin{abstract}
   In order to study the distribution of gas and ionizing radiation 
   around $\eta$ Car and their implications for its likely companion 
   star, we have examined high-excitation emission lines of [\ion{Ne}{3}], 
   [\ion{Fe}{3}], etc., in spectra obtained with the HST/STIS instrument 
   during 1998--2004.  Our principal results, some of them unexpected, 
   are:  (1) The high-excitation fluxes varied systematically and 
   non-trivially throughout $\eta$ Car's 5.5-year spectroscopic cycle.  
   Instead of rising to a plateau after the 1998 ``event,'' they changed 
   continuously with a maximum in mid-cycle.  (2) At one significant 
   location a brief, strong secondary maximum occurred just before the 
   2003.5 spectroscopic event.  (3) These emission lines are strongly 
   concentrated at  the ``Weigelt knots'' several hundred AU northwest 
   of the star.  With less certainty, [\ion{Ne}{3}] appears to be 
   somewhat more concentrated than [\ion{Fe}{3}]. 
   (4) A faster, blue-shifted component of each feature  appears 
   concentrated near the star and elongated perpendicular to the 
   system's bipolar axis.  This structure may be related to the 
   equatorial outflow and/or to dense material known to exist along 
   our line of sight to the star.  (5)  Using the photoionization program   
   Cloudy, we estimated the range of parameters for the hot secondary 
   star that would give satisfactory high-excitation line ratios in the 
   ejecta.  $T_\mathrm{eff} \approx 39000$ K and 
   $L \sim 4 \times 10^5 \; L_\odot$, for example, would be satisfactory.  
   The allowed region in parameter space is wider (and mostly less   
   luminous) than some previous authors suggested.  
\end{abstract}



\section{Introduction}

Eta Carinae is by far the most observable very massive star 
($M > 100 \, M_{\odot}$), the only supernova impostor that can be observed 
in great detail, and it has other extraordinary attributes   
\citep[see various papers in][]{2005ASPC..332.....H}.   The star and 
its ejecta produce more than 2000 identified emission lines, mostly of 
complex singly ionized species such as Fe$^+$ and Ni$^+$ 
 \citep{1953ApJ...118..234G,1966ApJ...146..126A,1967MNRAS.135...51T,  
 1989ApJS...71..983V,1994ApJ...422..626H,2001PhDT.........1Z}.
Satisfactory analyses of these low-ionization features have not been 
feasible, because they depend on radiative excitation processes which 
are extremely intricate.   A few higher-ionization spectral 
lines, however, offer valuable clues because they are less complicated:    
[\ion{Ne}{3}], [\ion{Ar}{3}], [\ion{Fe}{3}],  
etc.  They originate in slow-moving gas that was ejected from 
$\eta$ Car 100 to 200 years ago.  For reasons explained below,
they help to constrain the nature of a hot companion star which is 
thought to exist, but which has not been observed directly.  They 
also give clues  to the nature and distribution of ejecta within 3000 AU 
of the central star.\footnote{    
  Throughout this paper, if we omit individual references for either 
  a minor detail or a well-known generality about $\eta$ Car,  see 
  \citet{2005ASPC..332.....H,2001ASPC..242.....G,1999ASPC..179.....M,  
  1997ARA&A..35....1D}, and numerous refs.\ cited therein.  }    

In this paper we use Hubble Space Telescope (HST) data to explore the 
variability and spatial distribution of the high-ionization 
lines.  Several notable results emerge, including maps of the emission,  
continuous evolution of the intensities through $\eta$ Car's 5.54-year 
spectroscopic cycle, and a suggestive blue-shifted component of each line.  
Some of these facts contradict models that previously seemed credible.  
We also estimate limits to the secondary star's temperature and luminosity, 
based on the high-ionization features via photoionization calculations.  
We find a considerably broader range of allowed values than those 
proposed  by \citet{2005ApJ...624..973V}.

\subsection{Background information and astrophysical motivations}
\label{ssec:background}   

Several distinct types of spectra occur within 1{\arcsec} of $\eta$ Car.  
(1) The stellar wind produces broad emission and absorption lines, 
several hundred km s$^{-1}$ wide. (2) Slow-moving ejecta at 
$r \gtrsim 0.15{\arcsec}$ (i.e., more than 300 AU from the star) 
emit features narrower than $\mathrm{50~km~s^{-1}}$.  
(3) The same ejecta contain dust which reflects the stellar-wind 
spectrum.  (4) Faster ejecta also show both intrinsic and reflected 
spectra, though these are relatively faint in the central region.   
The high-excitation features belong mainly to category 2, 
but their excitation mechanisms are quite different from the narrow 
low-ionization lines.  The latter are strongly influenced by radiative 
processes such as resonance absorption of a UV photon leading to the 
emission of two or more lower-energy photons.  The high-ionization 
lines, however, depend on the same processes that dominate in ordinary 
nebulae -- especially photoionization by a hot star.  This star is 
most likely the  companion object mentioned in 
{\S}\ref{ssec:cycle} below.

The high-ionization lines have not been spatially mapped prior to 
this paper.  Most of them are narrow and thus do not 
represent the 300--1000 km s$^{-1}$ stellar wind. 
As a comparison, let us recall what is known about the spatial 
distribution of the {\it low-excitation\/} features.  The first 
HST   spectra of $\eta$ Car showed that the narrow [\ion{Fe}{2}] lines 
originate in the vicinity of the Weigelt knots, a set of three 
condensations 0.1{\arcsec}--0.3{\arcsec} northwest of the star 
\citep{1995AJ....109.1784D,1997AJ....113..335D,1986A&A...163L...5W}.  
These knots move outward rather slowly ($v \sim 40$ km s$^{-1}$), 
probably near the  equatorial plane of the system 
\citep{1988A&A...203L..21H,1995RMxAC...2...11W,1997AJ....113..335D,
1999A&A...344..211Z,2004AJ....127.1052D,2004ApJ...605..405S}).  
Later, Space Telescope Imaging Spectrograph (HST/STIS) data confirmed 
that low-excitation lines collectively account for much of the 
integrated brightness in that region 
\citep[e.g.][]{1999ASPC..179..144G,2000A&A...361..977J,1999A&A...344..211Z}. 
STIS could not resolve an individual Weigelt knot,    but 
in cases where its slit crossed one of them, the spatial distribution 
of [\ion{Fe}{2}] brightness peaked at the knot's position.  Meanwhile 
many HST images showed the individual knots at various wavelengths, 
albeit not well resolved.  Combining all these facts, most authors 
agree that the narrow low-excitation emission lines are largely 
concentrated in the Weigelt knots.  The high-ionization features,
on the other hand, had little influence on the HST images and on 
Weigelt's original speckle interferometry.\footnote{
   \citet{2000AJ....120..920S} asserted that HST/WFPC2 images with filters 
   F631N and F658N showed [\ion{S}{3}] and [\ion{N}{2}] 
   emission, respectively, in the Weigelt knots.  However, 
   (1) the F658N images were overexposed in the region of interest here, 
   and also included an indeterminate amount of red-shifted H$\alpha$ 
   emission;  and (2) there was no reason to assume that [\ion{S}{3}] 
   dominated in the F631N filter, which had been chosen to sample the 
   continuum instead.  STIS data show that in fact [\ion{S}{3}] 
   {\lm}6314 contributes less than 20\% of the signal on the Weigelt 
   knots.  The WFPC2 F631N images show primarily the fluctuating 
   continuum rather than [\ion{S}{3}] emission. }
Given their very different ionization states and excitation mechanisms, 
it is not safe to assume that they coincide with the knots.  They might 
represent diffuse lower-density gas surrounding the knots.  This 
uncertainty inspired the main work reported below, {\S}\ref{sec:emission}.


\subsection{Connections with $\eta$ Car's 5.54-year cycle, and 
   the secondary star}  
\label{ssec:cycle}   

Observers noticed long ago that these high-excitation lines occasionally 
weaken or even disappear and then gradually return to their normal 
strength after a few months \citep{1953ApJ...118..234G, 
1967MNRAS.135...51T,1983MNRAS.203..385W, 1984A&A...137...79Z}.  
Zanella et al.\ interpreted such ``spectroscopic events'' as shell 
ejections that temporarily quash the ionizing UV flux.   Eventually enough 
events had been observed to show that they recur with a period of 5.54 years 
\citep{1996ApJ...460L..49D,1994MNRAS.270..364W,2008MNRAS.384.1649D}. The 2--10 keV X-rays 
also disappear on each occasion \citep{1999ApJ...524..983I,
1999ASPC..179..266I,2005AJ....129.2018C}.  Many authors have discussed 
these phenomena, generally concluding that (1) there is a companion star 
in a highly eccentric 5.54-year orbit; (2) the X-rays are formed by 
the two colliding stellar winds; and (3) a spectroscopic event 
occurs near periastron due to a mass ejection and/or a disturbance in 
the wind and/or some sort of eclipse by the primary wind.  The projected 
separation between stars is less 
than 30 AU $\sim$ 13 mas, unresolvable by HST, and the secondary star 
has not been directly detected (at least not in any unambiguous way).  
For details see many refs.\ in \citet{2005ASPC..332.....H}, 
\citet{2001ASPC..242.....G}, and \citet{1999ASPC..179.....M}, and the 
physical considerations listed by \citet{2006ApJ...640..474M}.

The simplest conventional theories do not predict much variation of the 
high-excitation lines between spectroscopic events.  In those models 
the two stars are usually more than 10 AU apart and the densest regions 
of the primary wind are usually on the far side of the secondary, from 
our point of view.  Therefore $\eta$ Car's spectrum at average times 
has received comparatively little attention.  In {\S}\ref{sec:emission}, 
though, we find interesting behavior between the 1998.0 and 2003.5 events.

The secondary star's quantitative parameters remain open to question.
The opaque wind of the primary star probably has a characteristic 
temperature below 20000 K \citep{2001ApJ...553..837H}, too cool to 
account for the high-ionization lines.  Its hypothetical companion, on 
the other hand, presumably must be a massive star in order to cause 
an event, but not as massive as the primary and therefore less evolved; 
so we expect it to be hot.  The X-ray spectrum implies a 3000 km s$^{-1}$ 
wind \citep{2002A&A...383..636P}, and such a fast wind almost always 
implies $T_\mathrm{eff} \gtrsim 37000$ K.  

The usefulness of a hot companion for exciting $\eta$ Car's high-ionization 
lines has been recognized for more than a decade 
\citep{1997NewA....2..387D,1999ASPC..179..304D}.  This is an appealing idea 
because maybe we can deduce information about the secondary star from the 
high-excitation spectral features, via photoionization modeling.   
Since that object has not been observed directly, the 3000 km s$^{-1}$ 
wind speed is almost our only other clue to its parameters. 
\citet{2005ApJ...624..973V}  suggested spectral type and luminosity class 
O7.5 I, but it is not clear that such an object would fulfill the 
requirements. We explore this question in {\S}\ref{sec:photoionization} below.


\subsection{Contents of this paper}
\label{ssec:plan}   

We restrict our analysis to a few well-understood emission lines 
that require ionizing photons with $h{\nu}~\mathrm{> 16~eV}$.  Using spectra 
obtained with the HST/STIS
thoughout an entire 5.54-year cycle, we examine the spatial 
distributions and temporal behavior of [\ion{Ne}{3}] {\lm}3870, 
[\ion{Fe}{3}] {\lm}{\lm}4659,4703, and the narrow component of 
\ion{He}{1} {\lm}6680.\footnote{
   Throughout this paper we quote vacuum 
wavelengths and heliocentric Doppler velocities.} 
Then we 
calculate photoionization models to match the relative intensities 
of [\ion{Ne}{3}], [\ion{Ar}{3}], \ion{He}{1}, and other lines that 
require helium-ionizing photons ($h{\nu}~\mathrm{> 25~eV}$).   

In {\S}\ref{sec:observation} we describe the HST/STIS data.    
The variability and distributions of the emission features are 
described in {\S}\ref{sec:emission}, while {\S}\ref{sec:blue} 
concerns an interesting and significant blue-shifted component 
seen in all of these lines.   In {\S}\ref{sec:photoionization} 
we describe our photoionization modeling  with implications for 
the secondary star, and {\S}\ref{sec:discussion} 
concludes with a summary and brief discussion of our results.


\section{Observations, spectral analysis, and mapping technique}
\label{sec:observation}    

Eta Car was observed with HST/STIS from 1998.0 to 2004.2,  
covering an entire 5.54-year cycle.  The first 
of these observations occurred on 1998 January~1, a few weeks after 
the most critical time in the 1998.0 spectroscopic event.  
The last of them were obtained 2004 March~6, about 8 months after the 
2003.5 event.  Most of the observations were concentrated in 
mid-2003. 

This instrument provided the highest spatial resolution ever attained 
in spectroscopy of {\ec}.  The STIS/CCD  had three large-format 
(1024$\times$1024 
pixel) detectors with 0.05{\arcsec} square pixels, covering a nominal 
52{\arcsec}$\times$52{\arcsec} square field of view and operating from below 
2000 {\AA} to about 10000 {\AA}.  We used spectra observed with the  
52{\arcsec}$\times$0.1{\arcsec} slit in combination with the G230MB, G430M, 
and G750M gratings which provided resolutions  $R \sim$ 5000 to 10000. 
The  observations include  a variety of slit positions and orientations,
with a concentration at position angle $332\degree$ where the star 
and Weigelt knots B and D all fell within the slit.\footnote{
   The star and the three main Weigelt knots are customarily labeled 
   A, B, C, and D, with relative positions described by 
   \citet{1986A&A...163L...5W,1988A&A...203L..21H,2004ApJ...605..405S}, 
   and marked in our Fig.\ \ref{fig:fig2}.   
   Their separations expand by about 1\% per year.}  
Altogether, a map of the slit locations resembles a quasi-random ensemble 
of intersecting lines (see the figures in {\S}\ref{sec:emission} below).

HST/STIS/CCD data from the HST Treasury Project public 
archive\footnote{http://etacar.umn.edu/} 
include several improvements 
over the normal STScI pipeline and standard CALSTIS reductions. Modeling 
and interpolating the pixels in a special way improved the effective 
resolution along the cross-dispersion axis, compared to 
most other published STIS/CCD data \citep{2006hstc.conf..247D}. In this 
process the data were rebinned so that one pixel in the reduced data 
corresponds to 0.5 original CCD pixel, about 0.025{\arcsec}.  Initial 
bad/hot pixel removal, wavelength calibration and flux calibration 
matched or exceeded the STScI pipeline and CALSTIS.  We extracted many 
local spectra centered at uniformly-spaced locations along the STIS slit;  
the interval between successive locations was 0.025{\arcsec} and each 
extraction was 0.1{\arcsec} wide along the slit.  \citep[Characteristics 
of the original data precluded narrower extractions, 
see][]{2006hstc.conf..247D}.  At $\eta$ Car's distance $D \approx 2300$ pc    
\citep{1997ARA&A..35....1D,2001AJ....121.1569D,2006ApJ...644.1151S}, 
0.1{\arcsec} corresponds to a projected size of about 230 AU.

We examined the spatial distribution and the temporal behavior of 
[\ion{Ne}{3}] \lm3870, [\ion{Fe}{3}] \lm4659 and \lm4703, and the 
narrow component of \ion{He}{1} \lm6680. These were selected for 
minimal contamination by adjacent emission or absorption features.   
([\ion{Ne}{3}] $\lambda$3870, for instance, is weaker than   
[\ion{Ne}{3}] $\lambda$3969 but the latter nearly coincides with 
bright hydrogen and helium features.)  
For the spatial maps we used data extending from November 1998, 
11 months after the 1998.0 event, to the beginning of June 2003, 
a few weeks before the 2003.5 event (Table \ref{tab:table1};  we shall 
commment on the last dates below).  We measured fluxes by integrating 
each line above 
the continuum, or in some cases above an underlying broad emission 
profile.  Each measurement represented a spatial interval of about 
0.1{\arcsec} along the slit, the extraction width mentioned above.   

The assembly of a spatial map for each emission feature was a 
non-trivial exercise.  Of course each STIS spectrogram sampled only 
the region covered by the slit, effectively a strip about 0.13{\arcsec} 
wide if we allow for HST's spatial resolution.  On each observation 
date only one or two slit positions had been used;  but these varied in 
orientation and some of them were offset from the star, so the whole  
ensemble sampled a useful fraction of the area within 0.5{\arcsec} 
of the star.   Now consider any one of the measured emission lines.
Subject to later verification, suppose that the feature's relative 
spatial distribution did not vary much -- i.e., we tentatively assume 
that maps at any two different times closely 
resemble  each other except for a normalization factor.  (This 
statement does not apply to observations during an event, 
which we did not attempt to use.)     Since the line's 
brightness varied through the spectroscopic cycle,  we renormalized 
the values measured in each spectrogram so that the 
feature in question always had an adjusted value of unity at the 
only place that was observed on every occasion:  the central star.  
Fortunately  some narrow-line gas exists along our line of sight 
to the star,  producing [\ion{Ne}{3}], [\ion{Fe}{3}], and \ion{He}{1} 
lines superimposed on its spectrum.   These reference 
measurements were less precise than those elsewhere, because the 
superimposed narrow emission features were weak compared to the 
underlying stellar spectrum.  For each offset slit position, we 
adjusted the renormalization factor to maximize the consistency at 
intersections with other slit positions.

The above procedure obviously depends on some insecure assumptions, but 
we can apply consistency tests.   Some of the slit positions passing 
through the star were close together or were used on more 
than one occasion, and each offset slit position intersected 
several of the others (see figures in {\S}\ref{sec:emission}). 
Consequently a substantial number of points were independently 
re-observed at well-separated times.  Note that each slit sample 
had only one adjustable parameter,  its renormalization factor.  
Based on comparisons of the points re-observed in separate spectrograms,
the overall self-consistency of each map turned out to be quite 
good, and we found no local inconsistency worse than the expected 
measurement errors.  
Therefore our assumptions were sufficiently valid.\footnote{
   These statements are true even for the data obtained a few weeks 
   before the 2003.5 event.  If more observations had 
   been made in the mid-cycle period 1999--2002, we would have cautiously 
   excluded the pre-event 2003 data; but fortunately the latter turn 
   out to be consistent.   On 2003 May 17 and June 2 the offset STIS 
   slit intersected several earlier slit locations, see Table 
   \ref{tab:table1} 
   and figures in {\S}\ref{sec:emission}.   After the renormalization 
   factor had been applied, line strengths matched reasonably well 
   at those intersections. }    
Incidentally, possible time-delay effects cannot strongly influence 
these maps, since the light-travel-time for 
0.5{\arcsec} is only about a week at the distance of $\eta$ Car, 
and the cooling time in a He$^+$ region is less than a 
month for electron densities $n_{e} > 10^5$ cm$^{-3}$.   

In this paper, ``phase'' in the 5.54-year cycle is that used in 
the $\eta$ Car HST Treasury Program archive:  
$P =$  2023.0 days, $t_0 =$ MJD 50814.0 = J1998.0 exactly.       
This zero point corresponds to phase $\approx$ 0.009 in a system 
proposed by \citet{2008MNRAS.384.1649D}.  Zero points based on specific 
phenomena in the spectroscopic events should be avoided for two reasons:  
The 1998.0--2003.5--2009.0 events differed in major respects, and, 
also, successive revisions of $t_0$ chosen in this way have made 
comparisons among papers more difficult.  All proposed orbit models 
are far too imprecise to be useful in this regard.  The Treasury Program 
phase definition, on the other hand, has been extant {\it without 
alteration\/} for a number of years and its period is consistent with 
observations.  Considering that the Treasury Program archive is the 
largest available set of data on $\eta$ Car, it is the obvious 
reference standard.   


\section{Behavior and spatial distribution of the emission}
\label{sec:emission}   

The different photon energies required for [\ion{Ne}{3}], [\ion{Fe}{3}],
and \ion{He}{1} emission allow us to probe a range of physical parameters.
\ion{He}{1} $\lambda$6680 is a recombination line formed in a He$^+$ region, 
while [\ion{Ne}{3}] and [\ion{Fe}{3}] arise in Ne$^{++}$ and Fe$^{++}$ 
zones.  Ionization potentials of H$^0$, Fe$^+$, He$^0$, Ar$^+$,  and 
Ne$^+$ are 13.6, 16.2, 24.6, 27.6, and 41.0 eV respectively.  Thus, for 
reasons explained by \citet{2006agna.book.....O}, Fe$^{++}$ occurs mainly
in regions of H$^+$ and He$^0$ while Ne$^{++}$ and Ar$^{++}$ usually 
coexist with He$^+$.

\subsection{[\ion{Ne}{3}] emission line at 3870 \AA 
\label{sec:h1}}   

The top panel of Fig.\ \ref{fig:fig1} shows the time-dependent flux of the 
narrow [\ion{Ne}{3}] \lm3870  
emission line that was seen superimposed on STIS spectra of the star 
(at resolution $\sim$ 0.1{\arcsec})  through the 
spectroscopic cycle 1998--2003.  After a minimum lasting several 
months, the line intensity increased slowly until mid-cycle and then 
gradually decreased again. The intensity curve did {\it not\/} form a 
plateau.  Then, a few months before the 2003.5 event, the line 
intensity rapidly increased again before 
falling off toward its minimum.  This late-cycle rise 
resembled the behavior of the \ion{He}{2} emission and the observed 
X-ray light curve \citep{1999ASPC..179..266I,2006ApJ...640..474M};  perhaps 
the high-excitation emission at that time was somehow related to the 
X-rays or to the colliding-wind shocked zones (see {\S}6 in 
\citet{2008AJ....135.1249H} for reasons why this might occur).

Some, but not all, of this behavior pattern was noted by 
\citet{2008MNRAS.386.2330D} for [\ion{Ar}{3}] $\lambda$7138.     
They measured equivalent widths (not fluxes) in ground-based spectra 
including the star plus ejecta out to $r \sim 1{\arcsec}$.   
Their Fig.\ 1 differs from ours in two interesting respects:  
(1) The declines after mid-cycle do not match.   At phase 0.8, for 
instance, the high-excitation lines had already fallen by 50\% in the 
STIS results but less than 20\% in the ground-based data.    
(2) Damineli et al.\ did not observe a brief, dramatic flux 
increase around phase 0.9.  Most likely these discrepancies resulted 
from the very different spatial resolutions.  The ground-based spectra 
include an amorphous unresolved mixture of emission regions, whereas  
our Fig.\ \ref{fig:fig1} refers to a well-defined 0.1{\arcsec} locale 
which is particularly significant for a reason noted below.

A simple model would have predicted an extended plateau in Fig.\ 
\ref{fig:fig1}.  The [\ion{Ne}{3}] flux shown there is superimposed on 
the star and has a negative Doppler velocity of about $-40$ km s$^{-1}$;
so it must represent slow-moving gas located approximately between us 
and the star.  If it was ejected 50--200 years ago, this gas has now moved 
several hundred AU from the star.  If the orbit orientation is within the    
range favored by most authors
\citep{2001ASPC..242...53I,2008MNRAS.388L..39O}, then during most of 
the 5.5-year cycle the initial part of a path from the hot secondary 
star to the Ne$^{++}$ in question (i.e., toward us) should pass 
through the low-density secondary wind facing away from the 
primary star;  see Fig.\ 2 in \citet{2002A&A...383..636P}.  
We omit quantitative details here, but previous data would have 
allowed a scenario in which the ionizing photons  are not 
seriously depleted along such a path until they reach the 
observed Ne$^{++}$ region.  In that case the [\ion{Ne}{3}] brightness 
superimposed on the star would change little during most of the cycle.  
Fig.\ \ref{fig:fig1} contradicts this simple model.     

One can easily imagine qualitative explanations for the observed 
behavior after seeing it, but choosing the right one is harder.  
Most likely the ionizing photon path mentioned above {\it does\/} 
intersect a substantial and varying part of the dense primary wind.   
\citet{2009MNRAS.396.1308G} and \citet{2009MNRAS.394.1758P}, for instance, 
have discussed possible ``conventional'' models for the flow of dense 
gas outward from the binary system, allowing a substantial column density 
in the relevant sense.  Soker (priv.\ comm.), on the other hand, 
remarks that Fig.\ \ref{fig:fig1} is qualitatively consistent with 
a very different picture,  in which the secondary star is usually 
on the far rather than the near side of the primary  
\citep{2008MNRAS.390.1751K,2009MNRAS.394..923K,2009MNRAS.397.1426K}.  
A discussion of these issues is beyond the scope of this paper, but   
we note two points:  (1) Every proposed model depends on a number of 
assumptions which are not easy to verify, and (2) the information in 
Fig.\ \ref{fig:fig1} is essential for any realistic view of the problem.  
Instead of an ill-defined average of ejecta around the star, this figure
represents emission between us and the star at high spatial resolution.

Fig. \ref{fig:fig2} shows a spatial map of the narrow [\ion{Ne}{3}] \lm3870 
feature in slit positions observed with STIS.  In order to compensate for 
time variations, we renormalized the measurements so that the value is 
always unity at the position of the star (see {\S}\ref{sec:observation}). 
Most of the flux originates in the region of the Weigelt knots, at 
distances 0.1{\arcsec}--0.3{\arcsec} from the star.  The positions 
of the rather amorphous knots BCD are marked in the figure.     STIS could 
only marginally resolve them individually, but the peak obviously 
corresponds to Weigelt knot C.  In our map we can locate this peak at 
$r \approx 0.21{\arcsec}$, position angle $\approx$ 300{\degree}.  This corresponds 
well with measurements of knot C in HST images which gave 
$r \approx 0.22{\arcsec}$ in 1999--2003 (e.g.\ \citet{2004ApJ...605..405S}).

Higher flux in [\ion{Ne}{3}] \lm3870 was also observed at the locations 
of Weigelt knots B and D at position angle $\sim$ 335{\degree}.  Their 
peak intensities are roughly half that of knot C.  However, it is not 
possible to distinguish between B and D based only on Fig. \ref{fig:fig2}. 

Given the spatial-resolution limits of all HST data, the identification 
of [\ion{Ne}{3}] with the Weigelt knots in Fig.\ \ref{fig:fig2} is 
practically as good as the evidence for their association with [\ion{Fe}{2}] 
reviewed in {\S}\ref{ssec:background} above.  Evidently the narrow-line  
[\ion{Ne}{3}] emission does not chiefly represent a diffuse halo 
enveloping all the Weigelt knots, as seemed possible before.

\subsection{[\ion{Fe}{3}] emission lines at 4659 \AA\ and 4703 \AA\ 
\label{sec:h2}}    

We examined both [\ion{Fe}{3}] \lm4659 and \lm4703 in order to compare 
them for mutual consistency.  Their theoretical intensity ratio is 
1.85 \citep{1996A&AS..119..509N,1996A&AS..116..573Q}.   
Figs.\ \ref{fig:fig3} and \ref{fig:fig4} show maps of the narrow 
[\ion{Fe}{3}] \lm4659 and \lm4703 fluxes respectively.  In the same 
manner as for [\ion{Ne}{3}], we rescaled the measured [\ion{Fe}{3}] 
fluxes so the net value at the location of the star was always unity.   
Before rescaling, our measured flux of [\ion{Fe}{3}] \lm4659 at that  
position was about 1.9 times as high as for [\ion{Fe}{3}] \lm4703 
in each observation, close to the theoretical value.  
The two [\ion{Fe}{3}] maps are in 
excellent agreement with each other.   [\ion{Fe}{3}] emission originates 
in a slightly larger region than [\ion{Ne}{3}]. This is not very 
surprising, given the difference in ionization potentials; 
[\ion{Ne}{3}] occurs where helium is singly ionized but [\ion{Fe}{3}] 
originates in zones where hydrogen is ionized but helium is not.

Incidentally, just as this paper was completed we noticed a fainter 
additional [\ion{Fe}{3}] condensation about 0.48{\arcsec} NNE of the
star, in data obtained in June 2009 as part of a STIS re-commissioning 
exercise.\footnote{
     HST program 11506:  K.\ S.\ Noll, B.\ E.\ Woodgate, 
       C.\ R.\ Proffitt, \& T.\ R.\ Gull.}
We mention this knot because it has a {\it positive\/} Doppler velocity,  
about $+27$ km s$^{-1}$ rather than the values around $-40$ km s$^{-1}$
seen in BCD.   Its location was not sampled by the slit positions 
used in 1998--2004.   

In addition to [\ion{Ne}{3}], Fig.\ \ref{fig:fig1} shows the temporal 
variation of [\ion{Fe}{3}] {\lm}{\lm}4659,4703 on the position of the star.  
They display the same qualitative behavior as [\ion{Ne}{3}] \lm3870.    

This paper is not directly concerned with low-excitation features,
  but we must acknowledge that some of the narrow [\ion{Fe}{2}] lines 
also  showed temporary minima at phases around 0.8 in Fig.\ 
\ref{fig:fig1}, qualitatively like [\ion{Ne}{3}] and [\ion{Fe}{3}].  
 They all showed  a peak at phase 0.9 and a brief minimum during the 2003 event. 
 The  [\ion{Fe}{2}] lines varied differently from each other, 
 but none of them showed a strong growth around phase 0.3 as [\ion{Ne}{3}] did.

\subsection{\ion{He}{1} emission line at 6680 \AA 
\label{sec:h3}}  

\ion{He}{1} \lm6680 is a recombination line formed in He$^+$ 
\citep{2006agna.book.....O}.  In a spectrum of $\eta$ Car, this 
feature usually consists of a broad component formed in the 
stellar wind, plus the narrow component from slow-moving ejecta.  
With the spectral resolution of STIS ($\sim$ 40 km s$^{-1}$), 
the underlying broad emission makes the narrow \ion{He}{1} 
line far more difficult to measure than [\ion{Ne}{3}] and 
[\ion{Fe}{3}].  Consequently it is not a clear enough indicator 
of the time dependence to include in Fig.\ \ref{fig:fig1}.  
Our \ion{He}{1} \lm6680 map (Fig.\ \ref{fig:fig5}) is thus of 
lower quality than the others.  There is no evident disagreement 
compared to [\ion{Ne}{3}].       


\section{Blueshifted components of the high-excitation lines}
\label{sec:blue}   

In addition to the narrow lines discussed above, [\ion{Fe}{3}] 
and [\ion{Ne}{3}] also show separate, broader components extending 
between Doppler velocities $-250$ km s$^{-1}$ and $-400$ km s$^{-1}$.  
These also exist for other high-excitation lines 
such as [\ion{Ar}{3}], [\ion{S}{3}], [\ion{Si}{3}], etc.  In each case 
the blue-shifted component is much wider than the narrow lines but  
much narrower than the normal stellar wind emission features.  
\citep[$\eta$ Car's brightest Balmer lines usually have 
FWHM $\approx$ 400 to 500 km s$^{-1}$; see, e.g.,][]{2005AJ....129..900D}

\citet{1953MNRAS.113..211T,1967MNRAS.135...51T} recognized [\ion{Fe}{3}] and 
[\ion{S}{2}] components as emission blue-shifted by about $-300$ km s$^{-1}$.
\citet{1966ApJ...146..126A} listed these and similar blue-shifted 
\ion{Fe}{2} and [\ion{Fe}{2}] lines.  \citet{1984A&A...137...79Z} noted 
that some of them disappeared along with the ordinary narrow 
high-excitation lines during an event, particularly including [\ion{Ne}{3}].  
They concluded that the blue-shifted emission arose at larger distances 
from the star; but the Weigelt knots and polar/equatorial morphology were 
not known at that time.  Comparing the 1992 spectroscopic event   
to the 1995 midcycle state, \citet{1998A&AS..133..299D} confirmed the 
Zanella et al. statements about variability.  \citet{2009MNRAS.396.1308G} 
recently showed STIS spectrograms (not tracings) of some of these 
features, but did not produce spatial maps like those presented below.

Quantitative measurement of the blue-shifted [\ion{Ne}{3}] \lm3870 is 
difficult due to nearby \ion{Si}{2} \lm3864  and \ion{Cr}{2} \lm3867 
\citep{2001PhDT.........1Z}. We therefore concentrated on the blue-shifted 
component of [\ion{Fe}{3}] \lm4659, centered at about 4653 \AA. 
To verify that it is [\ion{Fe}{3}] emission, we compared the profiles 
of [\ion{Fe}{3}] \lm4659 and \lm4703.  Fig. \ref{fig:fig6} 
shows that the blue-shifted feature is undoubtedly a component of each 
[\ion{Fe}{3}] line. This figure also illustrates the complexity of the 
line profiles, tracing various ejection features. There is probably 
some material with intermediate Doppler shifts.

The temporal evolution of the blue-shifted component follows 
the spectroscopic cycle in roughly the same way as the narrow 
lines.  The normal and the shifted components appear and disappear 
together.  Fig. \ref{fig:fig7} compares the [\ion{Fe}{3}] \lm4659 emission 
line profile in spectra of the star at different phases in the cycle. 
The solid curve refers to a phase of 0.59, close to the time 
when the narrow high-excitation emission was strongest. 
The dashed curve is the profile shortly after the 2003.5 event, at phase 
$\approx$ 1.0. The narrow component at 4659 \AA\ disappeared completely, 
and the broad blue component was diminished. 

Similar to {\S}\ref{sec:emission},  in Figs.\ \ref{fig:fig8} and 
\ref{fig:fig9} we map the flux of the blue-shifted component of 
[\ion{Fe}{3}] \lm4659.  Most of it originates in the inner 
0.1{\arcsec} region, less than 250 AU from the star.  The distribution 
is not as sharp as a stellar point-source would be.  {\it The 
emission region  is detectably elongated to the northeast and 
southwest,\/} perpendicular to the axis of the bipolar Homunculus 
ejecta-nebula.\footnote{    
  \citet{2009MNRAS.396.1308G} mentioned ``diffuse arcs'' WSW and ENE 
  of the star, but they meant velocity vs.\ position correlations in 
  spectrograms, not spatial arcs.  ``WSW'' and ``ENE'' referred to the 
  position angle of the STIS slit when those obervations were made.
  Although they used direction names such as NE and ENE, those authors 
  did not describe the quantitative elongation 
  shown in our Fig.\ \ref{fig:fig9}.  Their discussion focussed instead 
  on velocity-position-time effects that one can see directly in the 
  spectrograms. }     

Considering the proximity of the blue component to the position of 
the star, its size, shape, and orientation, we suspect that Fig.\ 
\ref{fig:fig9} traces the near side of a latitude-dependent structure 
in the outer stellar wind.  Its observed blueshift of about $-300$ 
km s$^{-1}$ is smaller than the wind speed seen at the bottoms of 
the P Cyg absorption features 
\citep[e.g.,][]{2003ApJ...586..432S,2005AJ....129..900D},
but these two quantities sample the gas in very different ways.     
Electron densities cannot greatly exceed  $10^7$ cm$^{-3}$ in 
the relevant gas, because higher values would collisionally 
de-excite [\ion{Fe}{3}], [\ion{Ne}{3}], and [\ion{Ar}{3}] too strongly 
({\S}5.3 below).    Assuming that $\eta$ Car's mass-loss rate has the 
conventional value of $\sim 10^{-3} \; M_{\odot}$ yr$^{-1}$ and that 
its equatorial wind is less dense than its polar wind by a factor of 
order 3 \citep{2003ApJ...586..432S}, $n_e$   may fall below 
$10^7$ cm$^{-3}$ at {\it equatorial\/} radii 
$r \gtrsim 100 \; \mathrm{AU} \sim 0.05{\arcsec}$.  
At higher latitudes the corresponding radius would be larger.
Since these size scales are much larger than the 5.54-year orbit, very 
likely the dense, biconical inner polar wind ($r < 100$ AU) 
shields the outer, lower-density {\it high-latitude\/} zones from 
the secondary star's ionizing radiation.   In other words, at high latitudes 
the ionizing radiation is probably absorbed by gas that is too dense 
to produce strong forbidden lines.  Thus one expects the high-excitation
forbidden lines to be concentrated toward latitudes less than, say,
50{\degree} or so -- resulting in an oblate emission region.  If this 
picture is valid, evidently red-shifted emission from the far side cannot 
penetrate through the configuration.   Local inhomogeneities or 
``clumping'' do not alter the basic reasoning. 
\citet{2009MNRAS.396.1308G} have recently discussed relevant 
time-dependent structures at smaller size scales. Their models depend 
on a number of assumptions and do not make use of a polar wind; but in 
general they have little effect on our comments above.  Arguably the 
most significant well-defined observational clue is the elongation 
shown in Fig.\ \ref{fig:fig9} -- a new result.     

On the other hand, we can imagine two other locations for  
emission farther from the star.  First, the $-300$ km s$^{-1}$ Doppler 
shift is well-matched to the region where our line of sight to the star 
intersects the inner parts of the bipolar Homunculus ejecta-nebula and/or 
the ``Little Homunculus'' \citep{2003AJ....125.3222I}.  The main objection 
is that one might then expect the emission zone to have a projected 
size of 0.5{\arcsec} or larger.  The other possibility involves 
a remarkable observed fact: our line of sight to the star has about two 
magnitudes more extinction than our view of the Weigelt knots 
\citep{1995AJ....109.1784D,1999ASPC..179..116H,2001ApJ...553..837H}.  Dust 
grains cannot exist much closer to $\eta$ Car than $r \sim 150$ AU    
\citep{1997ARA&A..35....1D}, but apparently some very localized dusty 
material lies between us and the star.  Can its far side or inner side
be the region where the blue-shifted emission lines originate?  
Both the size scale in Fig.\ \ref{fig:fig9} and the Doppler velocity 
seem reasonable.  If this 
is the explanation, then the blue-shifted line components must be 
{\it intrinsically\/} much brighter than the observable narrow lines but are 
obscured by the dust.  (The same statement also applies to emission 
in the outer wind.  This model and the outer-wind idea mentioned 
earlier are not necessarily distinct from each other.)
  

\section{Photoionization modeling and the Nature of the Secondary Star }  
\label{sec:photoionization}  

As mentioned in {\S}1, the emission lines discussed here probably 
indicate photoionization by a hot star.   Consider, for 
instance, [\ion{Ne}{3}] $\lambda$3870.  This is a familiar feature in 
moderately high-excitation photoionized nebulae, where Ne$^{++}$ typically  
exists in a He$^+$ zone maintained by ionizing photons with 
$\mathrm{25~eV <}~h\nu~\mathrm{< 54~eV}$ \citep{2006agna.book.....O}.  Within that zone, 
[\ion{Ne}{3}] emission is collisionally excited by thermal electrons.
Suppose we attempt to construct a non-photoionized model for the ejecta 
of $\eta$ Car.  Since Ne$^+$ has an ionization potential 
of 41 eV, the creation of sufficient Ne$^{++}$ by thermal {\it collisional\/} 
ionization would require some mechanism -- e.g.,  
low-speed shock waves -- that can heat a substantial amount of material 
to $T \sim$ 20000--30000 K.  This is difficult to achieve
because a single low-speed shock carries insufficient energy;  and 
no such mechanism has been suggested by other data.\footnote{    
    Shock fronts with $v \sim 30$ km s$^{-1}$ can produce the required 
    temperature but the gas then cools rapidly. 
    With densities $n_e \sim 10^5$ to $10^8$ cm$^{-3}$ relevant in this 
    paper, insufficient material would be hot unless it is shocked 
    repeatedly, with a time interval less than a few months if  
    $n_e \sim 10^5$ cm$^{-3}$ or a few hours if $n_e \sim 10^8$ cm$^{-3}$.  
    A similar argument applies to heating 
    by MHD waves.  Strong shocks faster than 300 km s$^{-1}$ are known 
    to exist near $\eta$ Car but they  ionize neon far beyond Ne$^{++}$.  
    Other considerations and quantitative details are too lengthy to 
    discuss here. }
Moreover, as \citet{1984A&A...137...79Z} emphasized,   
[\ion{Ne}{3}] behaves very differently from the low-excitation lines 
during each spectroscopic event.  Thus, as a very probable working 
hypothesis,  {\it we assume that [\ion{Ne}{3}] $\lambda$3870  signals 
ordinary, relatively straightforward quasi-nebular photoionization.}  
So far as we know, no author has argued against this assumption
in the case of $\eta$ Car.  If correct then it is most likely 
true for [\ion{Ar}{3}] and [\ion{Fe}{3}] as well.  The low-excitation 
features, by contrast, are strongly influenced by 
local absorption in various UV permitted lines -- ``radiative pumping'' --  
which depends critically (and differently for each transition) on 
local velocity dispersions and density gradients.  In other words 
the high-excitation lines are theoretically more tractable than 
\ion{Fe}{2}, [\ion{Fe}{2}], and other low-excitation features.

But where do the Ne$^+$-ionizing photons above 41 eV come from?   
Conventional models for the primary stellar wind are too cool 
\citep{2001ApJ...553..837H}.   The secondary star, 
however, is expected to have a high effective temperature.   Its likely 
ability to create He$^+$ and Ne$^{++}$  in the ejecta was recognized 
as soon as its existence became probable \citep{1997NewA....2..387D,1999ASPC..179..304D}.  
Since Fe$^{++}$,  He$^+$, Ar$^{++}$, and Ne$^{++}$ depend on 
stellar photon fluxes just above 16, 25, 27, and 41 eV, 
the relative strengths of high-excitation 
lines may indicate the energy distribution of the star's 
ionizing UV photons, and thereby its temperature.  Since the glare of 
the primary star makes this object impossible to study directly, 
practically the only other quantitative information on it 
comes from the X-ray spectrum and from evolutionary constraints.   
Therefore, here we employ quasi-nebular photoionization 
calculations to model the relative high-excitation line strengths.
Differences from earlier calculations described by 
\citet{2005ApJ...624..973V} will be noted in {\S}\ref{ssec:diffs}.

\subsection{Procedure and assumptions}      
\label{ssec:Cloudy}  

We used version 08.00 of the photoionization program Cloudy 
\citep{1998PASP...110..761F} to simulate conditions observed at 
Weigelt knot C.  Since the primary star can safely be neglected 
in the high-ionization zones,  we assumed a simple configuration with 
one stellar source and a uniform-density cloud 
at distance $r \approx 10^{16} \, \mathrm{cm} \, \approx 700$ AU which 
is reasonable for knot C.   We chose a covering factor of 0.05 and 
a filling factor of unity for the gas.  We neglected possible 
UV extinction by dust within the He$^+$ region for three reasons: 
(1) local reddening of the [\ion{Fe}{2}] emission is known to be 
small in the Weigelt knots \citep{1999ASPC..179..116H,1995AJ....109.1784D}   
while the high-ionization zones should have smaller column densities;  
(2) dust grains associated with $\eta$ Car tend to be abnormally large 
\citep{1997ARA&A..35....1D} and therefore relatively ineffective  
at $h{\nu}~\mathrm{> 20~eV}$; and (3) not enough information is available 
to realistically include the effects of local dust on the ionizing 
radiation.  Altogether our assumptions represent an idealized view, but 
fortunately the line ratios used here depend only weakly on density 
and on geometrical details unless $n_{e}$ exceeds 
$10^7$ cm$^{-3}$.  A higher-density regime appears unlikely because 
collisional de-excitation would then suppress the forbidden lines 
too much.   If $n_e \lesssim 10^7$ cm$^{-3}$, the results depend 
mainly on the star's effective temperature $T_\mathrm{eff}$ and 
the ionization parameter $U_\mathrm{H}$, see {\S}\ref{ssec:star2} below.      

We assumed that the He$^+$ zone is ionization-limited, i.e., that 
it absorbs nearly all incident photons above 25 eV.  This assumption 
is based on three observed clues:
(1) lower-ionization regions clearly exist in and around the Weigelt knots, 
(2) almost no \ion{He}{1}, [\ion{Ne}{3}], and [\ion{Fe}{3}] emission is 
seen at radii outside the locations of the knots, and (3) ratios of the 
\ion{He}{1}, [\ion{Ne}{3}], and [\ion{Fe}{3}] lines do not vary much.  

Studies of $\eta$ Car's outer ejecta show a peculiar chemical composition 
due to the CNO cycle.  The gas is helium- and nitrogen-rich while carbon 
and oxygen are scarce
\citep{1986ApJ...305..867D,1999ASPC..179..134D}.\footnote{  
   The stellar wind spectrum appears consistent with the outer ejecta  
   but does not indicate abundances as well \citep{2001ApJ...553..837H}. } 
We considered two sets of abundances, ``compositions A and B'' 
which differ by a factor of two for elements heavier than helium  
(Table \ref{tab:table2}).  The mass fraction of C+N+O (mostly N) 
is about 0.63\% in composition A and 1.3\% in B.  The former value 
resembles some crude observational estimates for $\eta$ Car 
\citep{1986ApJ...305..867D}, while the latter is somewhat larger than Solar or 
Galactic material.  The distinction between A and B turns out to  
affect the results only weakly, as explained in {\S}\ref{ssec:star2}.    
Note that carbon and oxygen are too scarce to play an appreciable role.

We chose Weigelt knot C for this analysis because the high-excitation 
emission is strongest and best-defined there, see Fig.\ \ref{fig:fig2}.  
Knot C was observed well with HST/STIS on 2003 February 13 
($\mathrm{phase = 0.92}$ in the 5.54-year cycle) and 2003 May 17 
($\mathrm{phase = 0.97}$). We used data from 2003 February 13, obtained 
with STIS gratings G230MB, G430M and G750M and with the slit passing  
through the star and knot C along position angle 303{\degree}.  On 
that date the high-excitation lines were still strong, whereas by 
May 17 they had declined seriously as the 2003.5 event approached.
We measured well-isolated narrow high-excitation  lines as target 
values for the photoionization models. The resulting equivalent widths 
and apparent intensities are listed in columns 4 and 5 of Table 
\ref{tab:table3}.  An error of $\pm$30\% in an individual line 
strength, or perhaps even worse, would have little effect on our 
conclusions --  partly  because the photoionization models are very 
sensitive to the star's  $T_\mathrm{eff}$, and partly because the 
He/Ne/Si/Ar abundance ratios are more uncertain than the observed line ratios.

We estimated the reddening from the [\ion{Ar}{3}] {\lm3110}/{\lm5193} 
and  [\ion{Ar}{3}]  {\lm7138}/{\lm7753} flux ratios. The intrinsic value 
of the former ratio, for instance, is reliably known because the 
\lm3110 and \lm5193 lines share the same upper level;  and the case 
of {\lm7138}/{\lm7753} is similar.  
Other reddening indicators are less trustworthy because  
they use more than one ion species or differing upper levels 
or other model-dependent factors.  Since no physical model is available 
for $\eta$ Car's anomalously large grains or for instrumental effects
(see below), we adopted a standard approximation for wavelength-dependent 
extinction:  $A_\lambda \approx  a + b/\lambda$.
For small amounts of reddening this choice of mathematical 
form is not critical, provided that coefficient $b$ is adjusted to give 
the right average slope from violet to red wavelengths.  Fitting it to 
the [\ion{Ar}{3}] measurements, we found surprisingly little reddening,  
$E_\mathrm{B-V} = A_\mathrm{B} - A_\mathrm{V} \approx 0.2$ magnitude.

This is only half the amount that the [\ion{Fe}{2}] lines seem to 
indicate \citep{1995AJ....109.1784D,1999ASPC..179..116H}, but  
instrumental effects related to high spatial resolution probably account 
for most of  the discrepancy.   The normal wavelength dependence of HST's 
spatial resolution caused the slit throughput to decrease toward longer wavelengths 
for a localized source, and other effects also occurred in STIS data 
\citep{2006hstc.conf..247D}.  
Since these complications had smooth wavelength dependences, they are 
implicitly included in the {\it effective\/} or {\it apparent\/} reddening 
deduced from  the [\ion{Ar}{3}] line ratios.  In other words, the  
$E_{B-V}$ value mentioned above was really the true interstellar 
and circumstellar value minus a correction for instrumental effects.  
Meanwhile it is also possible that the [\ion{Fe}{2}] method gives an 
overestimate of $E_\mathrm{B-V}$.  No matter which effect dominates, 
the corrected relative line strengths are automatically valid   
to sufficient accuracy because the [\ion{Ar}{3}] comparison method 
is based on known intrinsic ratios.   (As mentioned earlier, 
``sufficient accuracy'' in this context would be $\pm$30\% or allowably 
even worse.  Most of the values in Table \ref{tab:table3} are 
expected to be better than this.) 

Thus, based on the [\ion{Ar}{3}] lines, we adopted the following 
correction for net reddening:   
\begin{displaymath}  
   I_\mathrm{corr} \, = \, 0.105 \, \exp  \left( 
    \frac{3740 \, \mathrm{\AA}}{\lambda} \right) \, I_\mathrm{obs} \, . 
\end{displaymath}  
Here the constant factor, which has no effect on the line {\it ratios\/} 
used in our photoionization analysis, is adjusted to give [\ion{Ne}{3}] 
\lm3870 a corrected value of 1.00.  As a result the corrected line 
strengths $I_\mathrm{corr}$ are intrinsic values relative to this line.  
They are listed in the last column of Table \ref{tab:table3}.

\subsection{Stellar model atmospheres}
\label{ssec:atmos}   

We explored a multidimensional grid of photoionization models, varying the 
effective stellar temperature, ionization parameter, and gas density. 
Our goal was to constrain these properties by comparing the calculated
intensity ratios of high-excitation emission lines to the observed 
ones. We tried four theoretical stellar atmosphere grids available 
in Cloudy. 
The Atlas models are LTE, plane-parallel, hydrostatic atmospheres 
with turbulent velocity distribution $\mathrm{2~km~s^{-1}}$ 
\citep{2004astro.ph..5087C}. The CoStar O-type models are non-LTE, 
line-blanketed model atmospheres, including stellar winds 
\citep{1997A&A...322..598S}.  The Tlusty models are non-LTE, 
line-blanketed, plane-parallel, hydrostatic O-star SEDs 
\citep{2003ApJS..146..417L}. The WM-basic O-star grids represent
non-LTE, line-blanketed, wind-blanketed hot stars 
\citep{2001A&A...375..161P}.

Unfortunately these four types of theoretical models disagree with 
each other in their UV spectral energy distributions.
Fig. \ref{fig:fig10} shows the continuum of an O-type main sequence 
star with $T_\mathrm{eff} = 40000$ K and $L = 10^5 \, L_{\odot}$      
according to each model.  Their differences are worst at high photon 
energies, particularly above 40 eV.   Table \ref{tab:table4} lists 
the total luminosities and photon rates that can ionize H$^0$, He$^0$,
and Ne$^+$ in a comparable set of the four model types.  The CoStar 
models do not include line opacity and therefore overestimate the 
far-UV flux because photon line-absorption and subsequent re-emission 
at longer wavelengths is not taken into account. The lower flux in the 
He$^{0}$ continuum ($h\nu~\mathrm{> 25~eV}$) of the WM-basic model compared to 
the Atlas models is probably due to non-LTE effects producing deeper 
line cores in the blocking lines. The WM-basic code uses a consistent 
treatment of line blocking and blanketing \citep{2002MNRAS.337.1309S}. 
After experimenting with all four types, we adopted this type of 
model. 


\subsection{Resulting constraints on the secondary star}
\label{ssec:star2}    
  
The Weigelt knots encompass a wide range of densities 
\citep{1999ASPC..179..116H}. 
We therefore varied the hydrogen density from 10$^{5}$ to 10$^{7}$ cm$^{-3}$, 
and for each density we varied the ionization parameter $U_\mathrm{H}$
between 10$^{-2}$ and 10$^{2}$.  In Cloudy, $U_\mathrm{H}$ is defined  
as the dimensionless quantity 
  $Q_\mathrm{H} /  4  \pi  r^2_0 n_\mathrm{H} c$, 
where $Q_\mathrm{H}$ is the rate of hydrogen-ionizing photons 
($h\nu > 13.6$ eV) emitted by the source star,  $r_0$ is the distance 
from that source to the illuminated face of the cloud, 
$n_\mathrm{H}$ is the hydrogen density, and $c$ is the speed of
light.\footnote{ 
  The ionization parameter determines the sharpness of each 
  ionization front, the coexistence of differing ionization 
  stages,  the ionized column density, and other physical 
  attributes, as explained by \citet{1979RvMP...51..715D}.  }
For our purpose, of course, we care most about the photon supply 
above 25 eV rather than 13.6 eV.  Cloudy produces ionization-limited
plane-parallel models, obviously a crude approximation to the true 
geometry.

Based on these calculations, Table \ref{tab:table5} and Fig.\ \ref{fig:fig11} 
show the effective temperatures of WM-basic stellar atmospheres that 
give the observed strengths of \ion{Si}{3}] \lm1892, [\ion{Ar}{3}] \lm7138, 
and \ion{He}{1} \lm6680 relative to [\ion{Ne}{3}] \lm3870.  ([\ion{Fe}{3}] and [\ion{S}{3}] 
are satisfactory in the 
models favored below.)    For ionization parameters 
$\log U_\mathrm{H} \gtrsim +1$ the required stellar temperature 
is roughly constant, while smaller values of 
$\log U_\mathrm{H}$ require progressively higher $T_\mathrm{eff}$.  
Since these results involve a subtle blend of uncertainties, the 
shaded regions in Fig.\ \ref{fig:fig11} represent factor-of-4 ranges
for each line ratio.  The best match to the observed data set occurs 
with $\log U_\mathrm{H} < -1$ and $T_\mathrm{eff} \sim 40000$ K.    
Here are some relevant considerations:  
\begin{enumerate}
  \item It is not surprising that chemical compositions A and B 
    ({\S}\ref{ssec:Cloudy}) give similar results.  Roughly speaking, 
    the hydrogen and helium recombination lines indicate numbers of ionizing 
    photons that have been absorbed, while the heavier-element emission 
    lines account for much of the cooling and thus represent the total 
    energy in the absorbed photons 
    \citep{1979RvMP...51..715D,2006agna.book.....O}.  Therefore, if 
    we alter the overall abundance of heavy elements relative to H+He, 
    the equilibrium gas temperature automatically adjusts so that the ratio 
    of heavy-element emission lines to hydrogen and helium lines does not 
    change much.  This ratio depends chiefly on the slope of the ionizing 
    source's spectral energy distribution.    
  \item For $\log U_\mathrm{H} > -1$, \ion{He}{1} $\lambda$6680 indicates 
    generally lower values of $T_\mathrm{eff}$ than the other lines do. 
    Very likely this clue is evidence for $\log U_\mathrm{H} < -1$;  but 
    we also note that models which are density-limited, or limited by  
    internal dust, or convex, can produce stronger helium  
    lines relative to [\ion{Ne}{3}].     
  \item Since relative abundances of individual heavy elements are 
    quite uncertain in $\eta$ Car,  and the reddening correction is 
    uncertain for the UV \ion{Si}{3}] line, the systematic difference 
    between [\ion{Ar}{3}] and \ion{Si}{3}] in Fig.\ \ref{fig:fig11}  
    is not very alarming.    
  \item Although the high-ionization emission lines are easier to model 
    than low-ionization features, some special processes may occur.  
    For example \ion{Si}{3}] \lm1892 may be enhanced by 
    resonance-favored two-photon ionization of Si$^+$ 
    \citep{2006A&A...452..253J}.  Such effects are unusual, however, and 
    unlikely to alter our basic conclusions.  (If the \ion{Si}{3}] effect      
    is strong, it produces a discrepancy between \ion{He}{1} and 
    \ion{Si}{3}] in photoionization models.)  
  \item  Models with $n_e  \gtrsim 2 \times 10^7$ cm$^{-3}$ produce 
    insufficient [\ion{Ne}{3}] and [\ion{Ar}{3}] emission compared to 
    \ion{He}{1} and \ion{Si}{3}], regardless of the stellar temperature.   
    This is because the characteristic electron densities for 
    collisional de-excitation of [\ion{Ne}{3}] and [\ion{Ar}{3}] are 
    roughly $10^7$ and $5 \times 10^6$ cm$^{-3}$ respectively.
  \item Models with $n_e << 10^6$ cm$^{-3}$ are ruled 
    out by  geometrical considerations.  For reasons mentioned in  
    {\S}\ref{ssec:Cloudy}, the He$^+$ region cannot be  much smaller 
    than the ionization-limited thickness.  This characteristic 
    linear size $x$ is proportional to $U_\mathrm{H}/n_e$.  
    In Table \ref{tab:table5} and Fig.\ \ref{fig:fig11} we have 
    indicated models that are unsuitable because $x \, > \, 250$ AU, 
    i.e., an ionization-limited He$^+$ zone would be larger than the 
    region of the Weigelt knots.  This criterion excludes all our 
    models with $n_e = 10^5$ cm$^{-3}$, and those which have 
    $n_e = 10^6$ cm$^{-3}$ and $\log U_\mathrm{H} > -0.65$.   
  \item A truly realistic model would include a range or distribution 
    of gas densities, but we don't have enough observables to do this.    
    At present one can say only that the ``representative density'' 
    in a simplified model is very likely about half the maximum 
    density which exists in the real, non-uniform He$^+$ gas, in 
    an order-of-magnitude sense.  (This statement can be wrong 
    if the configuration is unexpectedly complex.)   
  \item For a given gas density $n_e$ and stellar temperature 
    $T_\mathrm{eff}$, one can estimate the luminosity $L$ that would 
    produce the assumed value of $U_\mathrm{H}$ in Weigelt knot C.  
    As shown in the last column of Table  \ref{tab:table5}, some choices 
    of ($n_e, \, U_\mathrm{H}$) lead to absurdly small or large 
    values of $L$.  Most important, the secondary star cannot 
    exceed about $10^6 \, L_{\odot}$ since its presence is not evident 
    in the UV spectra obtained with HST/STIS.  (The primary star has 
    $L \approx 4 \times 10^6 \, L_{\odot}$.)
\end{enumerate} 

The H-R diagram in Fig.\ \ref{fig:fig12} summarizes our conclusions 
about the secondary star, assisted by a few more clues.  Here two broad 
curves mark the results of our photoionization calculations with 
$n_e = 10^7$ cm$^{-3}$ and $10^6$ cm$^{-3}$;  the upper end of the 
latter curve is limited by criterion 6 above.  If we assume  that 
the two stars have the same age, then evolutionary circumstances further 
constrain the parameters.  The minimum age for the primary star 
to have become helium-rich is 0.5 Myr \citep{1999ASPC..179..367I},  but it 
is most likely 2--3 Myr old based on its instabilities and probable 
association with the cluster Trumpler 16 \citep{1995RMxAC...2...51W}.  
The lifetime of any star above $\sim \, 60 \, M_{\odot}$ is 
roughly 3 Myr.   Therefore, in Fig.\ \ref{fig:fig12} we show isochrones 
for 0.5, 2, and 3 Myr as well as some evolutionary tracks, 
all adapted from \citet{2005A&A...436.1049M}.  We can base yet another 
constraint on the observed X-ray spectrum, which indicates a 
secondary wind speed close to 3000 km s$^{-1}$ \citep{2002A&A...383..636P}.  
Practically all stars with winds that fast have $T_\mathrm{eff} > 37000$ K,
marked by a vertical dotted line in Fig.\ \ref{fig:fig12} 
\citep[][and references cited therein]{2000ARA&A..38..613K}. 

Altogether, then, {\it we expect the secondary star to lie somewhere 
within the shaded polygon in Fig.\ \ref{fig:fig12},\/} with 
the following limits:
(1) The right-hand boundary is the X-ray-implied minimum temperature.  
Even without this argument we would deduce a similar limit  
from criterion 6 listed above for our photoionization models. 
(2) The upper limit is $L \approx 10^6 \, L_{\odot}$, criterion 8. 
(3) The upper-left boundary is based on our photoionization criterion
5, $n_e \lesssim 2 \times 10^7$ cm$^{-3}$.  
(4) The lower-left boundary is the 0.5 Myr isochrone, and perhaps 
this limit should be moved rightward in the diagram to some age 
greater than 1 Myr.  A star with $M_{ZAMS} \sim 40$ to $60 \, M_{\odot}$ and 
$T_\mathrm{eff} \approx 40000$ K, for example, would satisfy all 
these requirements.  The corresponding spectral type would be O4 or O5 
\citep{2005A&A...436.1049M}.   If, however, the system is more than 
2 Myr old,  then the secondary's zero-age mass was probably less than 
$50 \, M_{\odot}$.  (All statements related to isochrones obviously 
depend on the evolution models, though.)

If the entire region around $\eta$ Car were clearly visible, then the 
absolute brightnesses of the high-excitation emission lines would 
indirectly indicate the secondary star's luminosity. 
In fact the local extinction is far too patchy, but we attempted a
crude estimate based on the {\it observable\/} [\ion{Ne}{3}] brightness.  
In our Cloudy calculations with reasonable input parameters, the ionizing 
star's luminosity was around 2300 times the luminosity of 
[\ion{Ne}{3}] {\lm}3870.  Measuring the total flux of this line in 
the Weigelt knots, and assuming reasonable factors for the extinction 
and the solid angle intercepted by the knots, we found 
$\log L/L_{\odot} \sim 5.5$.  This estimate is very rough, but  
it is consistent with Fig.\ \ref{fig:fig12}  and thereby suggests 
that the absolute fluxes are reasonable.

\citet{2002A&A...383..636P} assumed that the secondary star has a mass-loss 
rate close to $10^{-5} \; M_\odot$ yr$^{-1}$, in order to obtain 
sufficient X-ray luminosity in their colliding-wind models. Such 
a high rate would be very unusual, and not entirely consistent 
with our discussion above.  Possibly this is an argument in favor of 
a more luminous secondary star;  but on the other hand the mass-loss 
estimate is not very robust.  Only a small fraction of the wind's 
kinetic energy is converted to observable X-rays, via some complicated 
efficiency factors.  The secondary wind {\it speed,\/} however,  
almost directly determines the average temperature seen in the 2--10 keV 
X-ray spectrum.  This is why we consider the 3000 km s$^{-1}$ speed 
estimate to be relatively more trustworthy than the mass-loss rate.

One can imagine models that are seriously affected by changes in the 
secondary star.  For instance, conceivably it was originally more 
massive but has now become a Wolf-Rayet object.  
\citet{2007ApJ...661..490S} has proposed another scenario, 
wherein the secondary star accreted a large amount of 
mass during the great eruption 160-170 years ago, and has not 
yet returned to its normal thermal equilibrium.  (See also 
\citet{2009NewA...14...11K}.)  Apart from obvious complications   
and a lack of substantive evidence for them, models of these types 
are  clearly not the {\it simplest\/} possibilities.  They have 
multiple adjustable parameters, and thus are beyond the scope of 
this paper.   (Note, incidentally, that our photoionization 
calculations do not favor an extremely hot star with with 
$T_\mathrm{eff} > 45000$ K.)

\subsection{Comparison with previous calculations}
\label{ssec:diffs}  

\citet{2005ApJ...624..973V} reported an earlier set of photoionization 
calculations very much like ours, with the same goal;  but they 
deduced appreciably different parameters for the 
secondary star.  Like us, they used Cloudy with WM-basic atmospheres, 
uniform density in the ionized material, etc.  They used earlier 
STIS data, but the high-excitation line ratios were similar 
to ours within the uncertainties.  Verner et al.\ appear to have  
supposed that the high-ionization lines originate much closer to the 
star than the locations of the Weigelt knots, but this probably had 
little effect on their results.  They described photoionization models 
with specific sets of parameters, but did not show systematic maps 
of parameter space as in our Table \ref{tab:table5} and 
Figs.\ \ref{fig:fig11} and \ref{fig:fig12}.

There are three notable differences between their results and ours:  
The allowed region in Fig.\ \ref{fig:fig12} is much larger than their 
discussion seems to imply, their suggested {\it upper\/} limit 
for $T_\mathrm{eff}$ is barely above our {\it lower\/} 
limit, and they assigned a maximal luminosity to the secondary 
star.  Indeed Verner et al.\ proposed that it has 
$T_\mathrm{eff} \approx 37200$ K and $\log L/L_{\odot} \approx 5.97$, 
marked with a small ``V'' in our Fig.\ \ref{fig:fig12}.  (They classified 
it ``O7.5 I'',  but according to \citet{2005A&A...436.1049M} an O5.5 
supergiant would have that temperature.)  Such an object is not  
excluded by the calculations, but in our view it has three  
disadvantages:  It has the minimum temperature required for a 
3000 km s$^{-1}$ wind, it requires the age of the system to be less 
than 2 Myr, and it has practically the largest allowable luminosity. 
So far as we know, there is no strong argument against lower luminosities.  
In summary, given the limited information currently available 
({\S}\ref{ssec:star2} above), the parameters suggested by Verner 
et al.\ are not the most suitable choices for assessing the nature 
of $\eta$ Car's companion at this time.

\section{Summary}
\label{sec:discussion}   

Here we have reported the behavior and the spatial origins of $\eta$ Car's 
quasi-nebular high-excitation emission between spectroscopic 
events.  Neither of these questions was previously explored with adequate 
spatial and temporal sampling.  
Indeed the 1998--2004 HST/STIS spectra appear to constitute the only 
suitable existing data set.  Thus our main results were previously 
vague or unknown, and some of them contradict models that would otherwise 
have seemed credible.

\begin{figure}
\epsscale{1}
\plotone{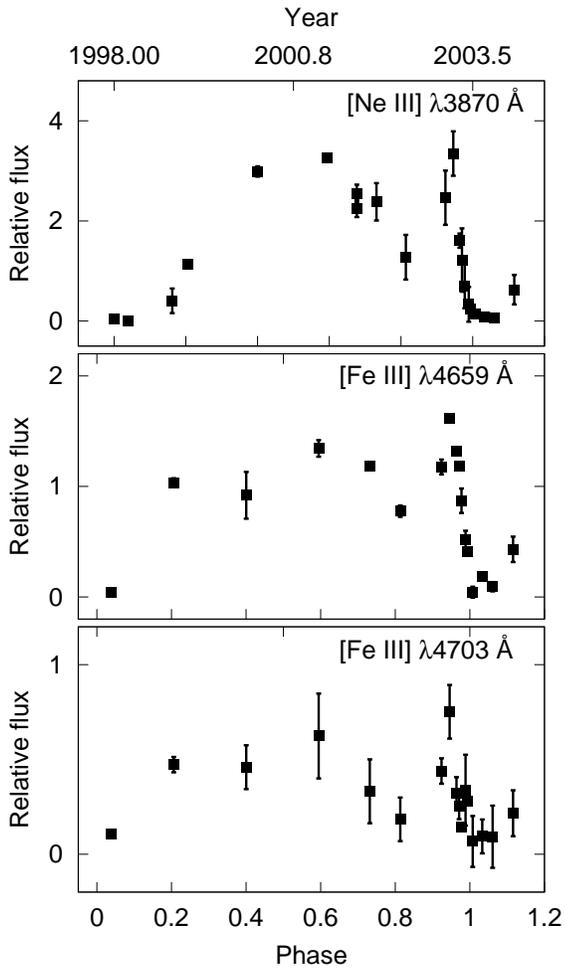}   
    \caption{Strengths of the narrow [\ion{Ne}{3}] \lm3870 and [\ion{Fe}{3}] 
    {\lm}{\lm}4659,4703 emission lines measured in HST/STIS spectra 
    of $\eta$ Car through its spectroscopic cycle.  These represent gas 
    along the line of sight to the star, seen with 0.1{\arcsec} spatial 
    resolution and not including the Weigelt knots.  Features in the 
    knots vary in a similar way.  \label{fig:fig1}}
\end{figure}

\begin{figure}
\plotone{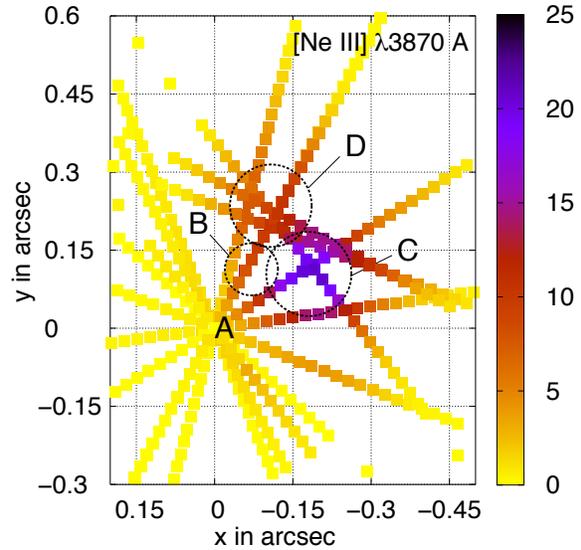}   
   \caption{Spatial map of the narrow [\ion{Ne}{3}] \lm3870 line.
    Position A marks the central star while B, C, and D are the 
    Weigelt knots.  Flux values are normalized so that this emission 
    line's net flux is unity at the star's position.  \label{fig:fig2}}
\end{figure}

\begin{figure}
\plotone{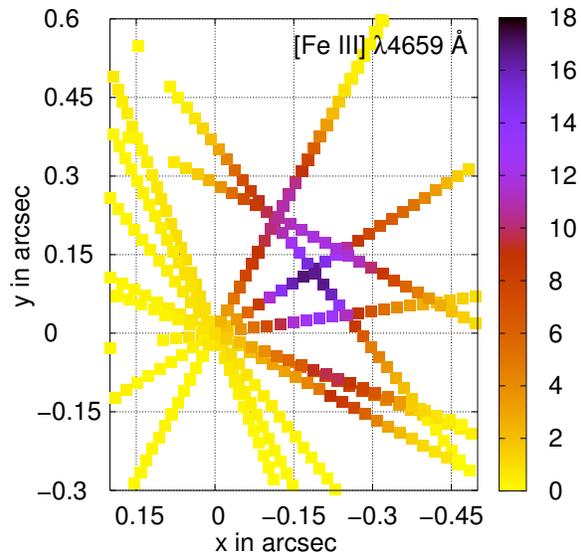}   
   \caption{Spatial map of the narrow [\ion{Fe}{3}] \lm 4659 line, 
   with the same format as Fig.\ \ref{fig:fig2}.  \label{fig:fig3}}
\end{figure}

\begin{figure}
\plotone{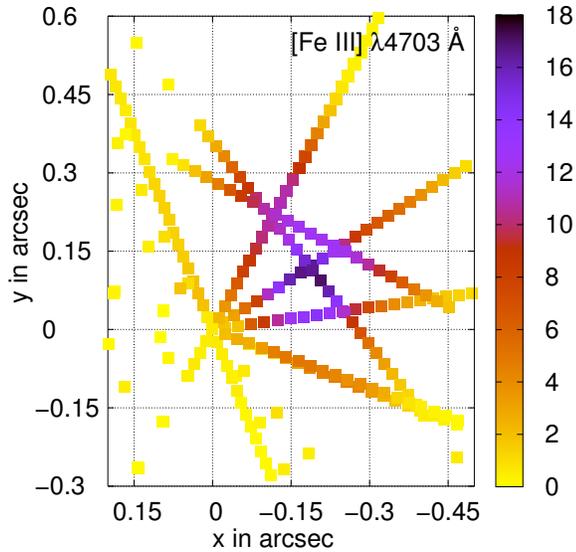}   
   \caption{Similar to Fig.\ \ref{fig:fig2}--\ref{fig:fig3}, but showing [\ion{Fe}{3} \lm 4703.
   This map is based on the same spectrograms as Fig.\ \ref{fig:fig3} 
   but the measurements were independent.  \label{fig:fig4}}
\end{figure}

\begin{figure}[!ht]
\plotone{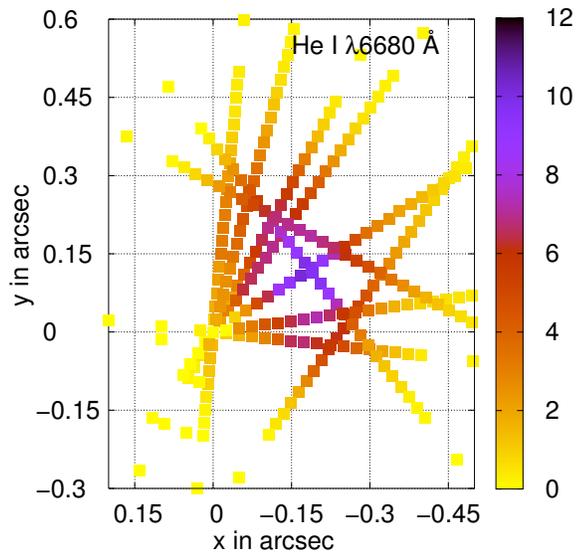}   
   \caption{Map of the the narrow \ion{He}{1} \lm6680 emission line, 
   shown in the same way as Figs.\ \ref{fig:fig2}--\ref{fig:fig4}.  
   \label{fig:fig5}}
\end{figure}

\begin{figure}
\plotone{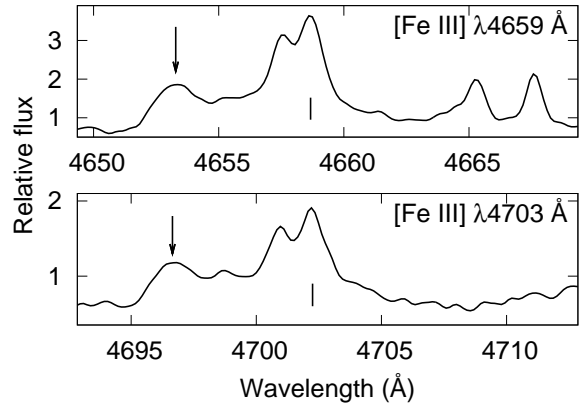}   
  \caption{Profiles of the two [\ion{Fe}{3}] emission lines that we 
  measured at the position of the star.  Arrows mark the broad 
  negative-velocity emission and vertical ticks show the 
  heliocentric zero-shift wavelengths. \label{fig:fig6}}
\end{figure}

\begin{figure}
\plotone{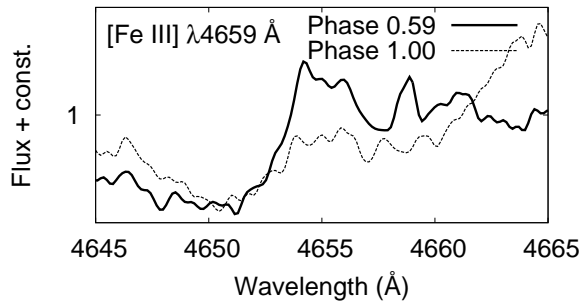}     
  \caption{Comparison of the [\ion{Fe}{3}] \lm4659 emission components 
  at two different phases in the 5.54-year cycle.   The solid line 
  represents a phase of 0.59 near high-excitation maximum, while the 
  dashed line shows phase 1.00 during the 2003.5 event. \label{fig:fig7}}
\end{figure}

\begin{figure}
\plotone{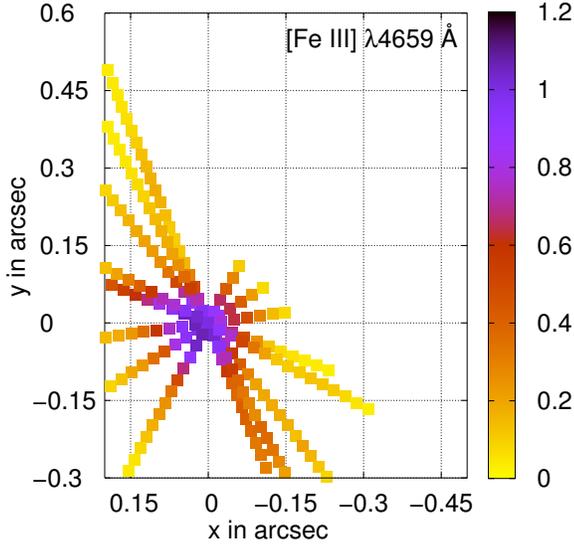}    
  \caption{Spatial map of the relative flux in the blue-shifted component 
  of [\ion{Fe}{3}] \lm4659.  The flux is normalized to unity at the 
  position of the star.  When measured carefully the bright region 
  is elongated along the NE--SW direction, see Fig.\ \ref{fig:fig9}. 
  \label{fig:fig8}}
\end{figure}

\begin{figure}
\plotone{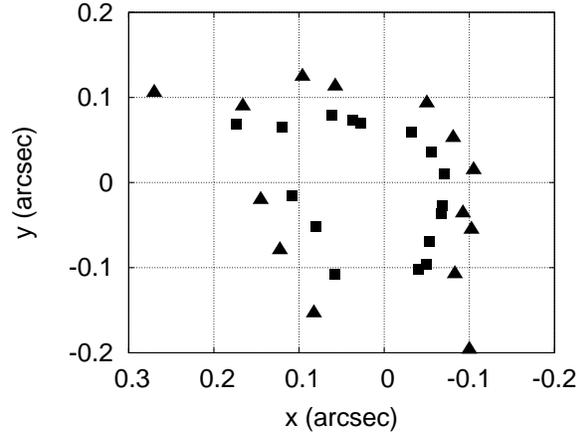}   
   \caption{Expanded view of intensity contours of the blue-shifted 
   [\ion{Fe}{3}] \lm4659, using the same measurements as 
   Fig.\ \ref{fig:fig8}.  Squares ($\blacksquare$) and triangles 
   ($\blacktriangle$) indicate values of 0.5 and 0.25 respectively, 
   relative to the value at the star's location. \label{fig:fig9}}
   The contours are significantly elongated NE--SW, i.e., toward
   the upper left and lower right.  
\end{figure}

\begin{figure}
\centering
\plotone{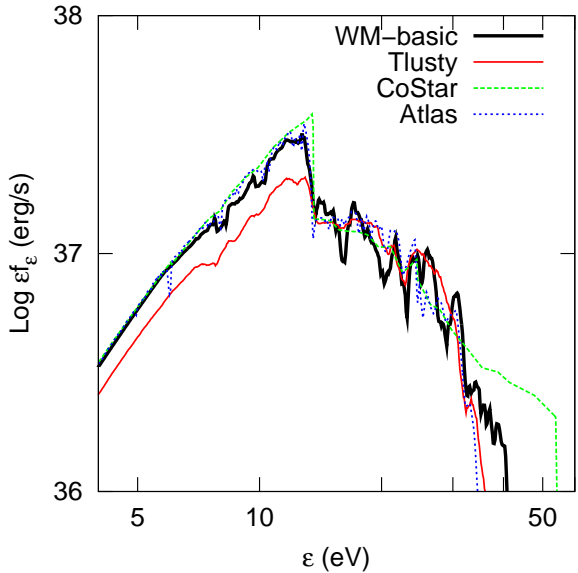}  
   \caption{Comparison of the stellar continuum in Atlas, CoStar, Tlusty 
   and WM-basic atmosphere models with $T_\mathrm{eff} = 40000$ K. Tlusty and WM-basic models were smoothed.
      \label{fig:fig10}}
\end{figure}

\begin{figure*}
\centering
\includegraphics[width=1\textwidth]{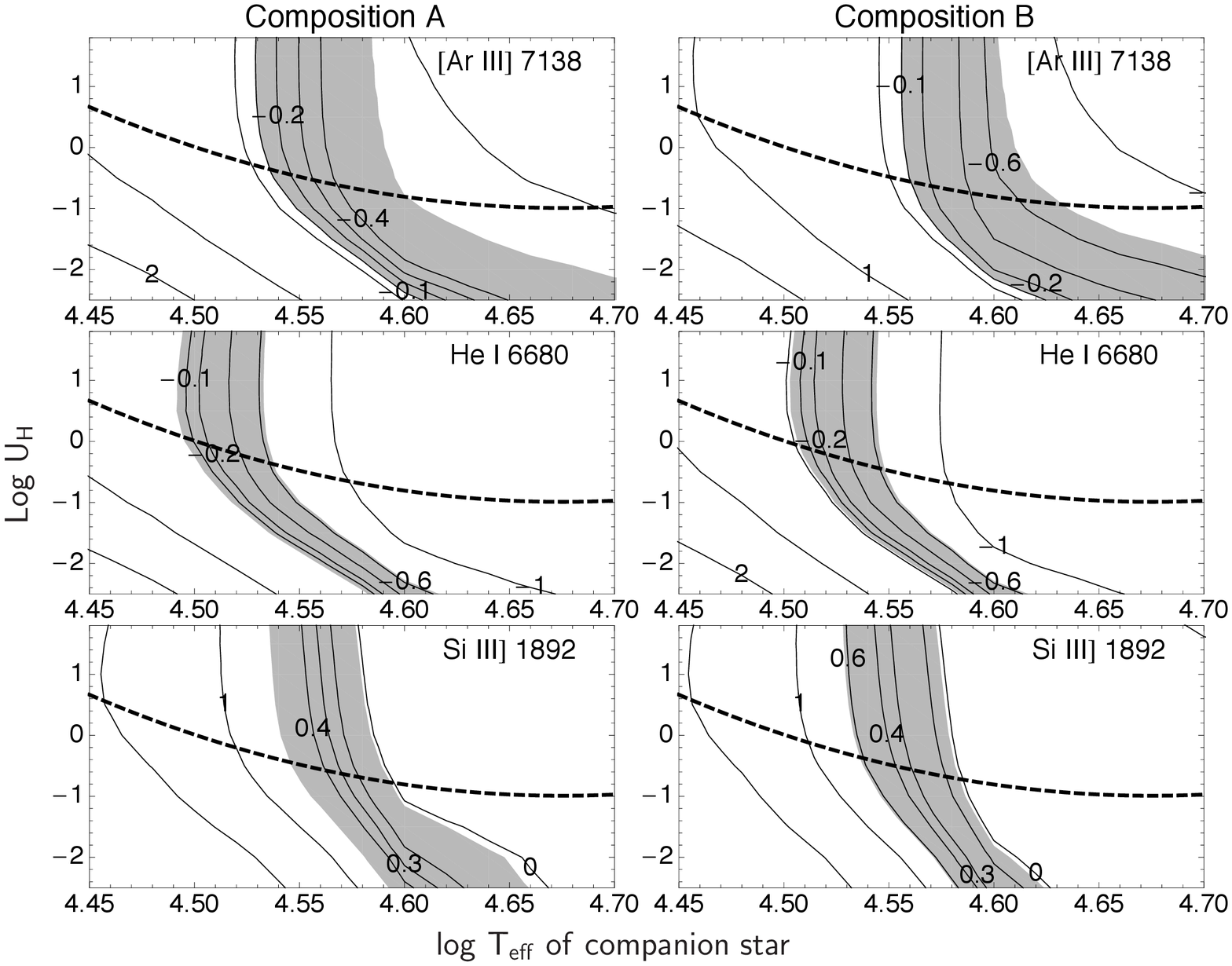}   
  \caption{Contour plots of line intensity ratios 
  [\ion{Ar}{3}] \lm7138/[\ion{Ne}{3}] \lm3870 (top),   
  \ion{He}{1} \lm6680/[\ion{Ne}{3}] \lm3870 (middle), and 
  \ion{Si}{3}] \lm1892/[\ion{Ne}{3}] \lm3870 (bottom) in our photoionization 
  models with $n_{\mathrm{H}}\mathrm{ = 10^{6}~cm^{-3}}$. Shaded areas 
  correspond to the range of line intensity ratio between 0.5 and 2 times 
  the observed value.  Left and right columns refer to chemical compositions 
  A and B respectively.  Models above the dashed curve are spatially 
  too extended, violating criterion 6 in the text.  \label{fig:fig11}  }
\end{figure*}

For instance, after the 1998 spectroscopic event the high-excitation 
features  did not simply recover to constant ``normal'' intensities, but 
instead {\it varied  systematically throughout the 5.5-year cycle\/}  
({\S}{\S}\ref{sec:h1}, \ref{sec:h2}).  [\ion{Ne}{3}] and [\ion{Fe}{3}]
evolved through four distinct stages (Fig.\ \ref{fig:fig1}):  
\begin{enumerate}
  \item  As expected based on previous events, 
    the weak or near-zero state persisted for several months.  This 
    was long enough for a companion star to sweep through more than 
    180{\degree} of longitude around periastron, if its orbit eccentricity 
    is at least 0.8 as seems likely.  
  \item Subsequent growth to near-maximum intensity was interestingly 
    slow, delaying the maximum until more than two years after the 
    event -- unlike what a simple model would predict.
  \item Then, instead of leveling off,  the intensity soon began to 
    decline, following a sort of parabolic trajectory in time.  
  \item A few months before the 2003.5 event, however, it briskly rose  
    to a second maximum just before declining again to near-zero.
\end{enumerate}    
We directly measured the double-peaked cycle in gas along the line 
of sight to the star, but at least stages 1--3 seem broadly valid 
in our data on the Weigelt knots BCD as well.  As noted in {\S}\ref{sec:h1},  
\citet{2008MNRAS.386.2330D} recognized those stages for one [\ion{Ar}{3}] 
line at much lower spatial resolution, but there is a quantitative 
disagreement in stage 3 and they did not see stage 4.  

If this time-pattern was caused by some complex variation of the 
circumstellar extinction, then in order to include BCD it must extend 
out to $r \gtrsim 800$ AU, which seems unlikely in terms of grain 
physics.  Apparently, then, during most of the 1998--2003 cycle 
the amounts of Ne$^{+2}$, Fe$^{+2}$, etc.\ 
in BCD were significantly less than their maxima.  Despite our comments 
at the beginning of {\S}\ref{sec:photoionization}, the 
secondary maximum might conceivably be related to the colliding-wind 
shock structure which was very strong in the months before an event.  
The softest X-rays may play a role then \citep{2006ApJ...640..474M}.   
A detailed speculative/theoretical interpretation of Fig.\ \ref{fig:fig1} 
would require calculations beyond the scope of this paper.  
\citet{2006PASP..118..697M,2009arXiv0908.1627M} have described 
other phenomena observed midway between $\eta$ Car's spectroscopic 
events, emphasizing that its mid-cycle behavior deserves more attention.

In {\S}3 we also showed spatial maps of the [\ion{Ne}{3}] and [\ion{Fe}{3}] 
emission in the sub-arcsecond vicinity of \ec.  Informal examinations   
of [\ion{Ar}{3}], [\ion{S}{3}] and [\ion{Si}{3}] gave similar impressions.    
Evidently {\it high-excitation emission is strongly concentrated in the 
Weigelt knots BCD\/} located at a distances  0.15{\arcsec}--0.35{\arcsec} 
northwest of the star, rather than in a surrounding halo or in the region 
between the star and the knots as seemed possible before.  Prior to this    
work most researchers would very likely have {\it assumed\/} that each knot 
produces high-excitation emission, but here we have given the first clear 
proof.  Weigelt blob C has the highest flux.  Maps of the various ion 
species all show basically the same picture, except that those with higher 
ionization potentials seem a little more compact.  Unfortunately it will 
be difficult to improve these maps in the future, because the rising 
brightness of the central star is rapidly making the knots harder 
to observe \citep{2006AJ....132.2717M}.

The high-excitation lines all have blue-shifted components in 
the velocity range $-250$ to $-400$ km~s$^{-1}$.  This fact was known 
before, but in {\S}4 we have shown that (1) it originates mostly 
within 0.15{\arcsec} of the central star, and (2) this region is 
elongated along the equatorial (not polar) axis of the $\eta$ Car 
system.  In {\S}4 we proposed several plausible interpretations involving 
the three-dimensional distribution of stellar ejecta.  These 
alternatives are not mutually exclusive.

\begin{figure}
\centering
\plotone{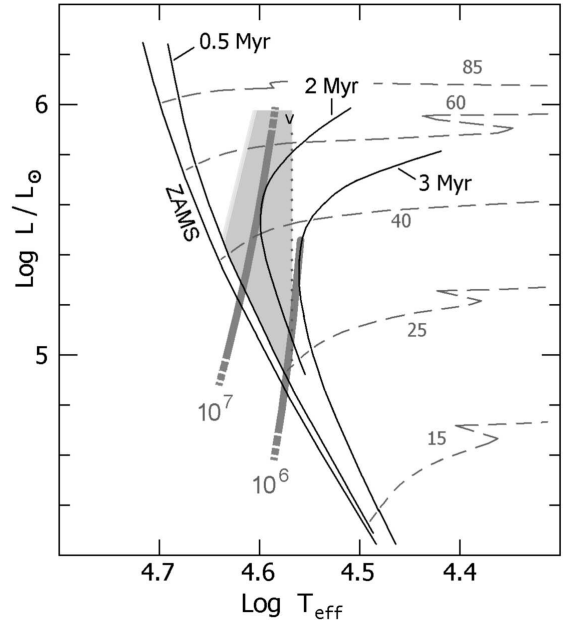}   
   \caption{Likely positions for $\eta$ Car's secondary star in the 
   H-R diagram.  Solid curves show isochrones for 0.5, 2, and 3 Myr, 
   while dashed curves show evolution tracks for initial masses 15 to 85 
   $M_\odot$, all adapted from \citet{2005A&A...436.1049M}.  The two 
   broad lines show $L$ vs.\ $T_\mathrm{eff}$ correlations in our 
   photoionization models with $n_\mathrm{H} = 10^7$ and $10^6$ cm$^{-3}$. 
   A small `V' indicates a model suggested by \citet{2005ApJ...624..973V}.  
   The shaded polygon is the region allowed by various considerations 
   (see text).  \label{fig:fig12}}
\end{figure}

Assuming that a hot companion star is responsible for the narrow high 
excitation emission lines,  in {\S}\ref{sec:photoionization} we found 
constraints on its parameters based on photoionization calculations. 
Our results are consistent with evolutionary considerations and, 
independently, with the colliding-wind X-ray temperature.  
Fig.\ \ref{fig:fig12} shows that the allowed region in parameter space is 
larger than some previous authors suggested.  For example, an O4--O6 giant 
with $L \sim 4 \times 10^5 \; L_\odot$, $T_\mathrm{eff} \approx 39000$ K, 
and $M_\mathrm{ZAMS} \sim$ 40--50 $M_\odot$ would fall near the center 
of the allowed range.   Our calculations did not exclude moderately 
higher masses and luminosities, but neither did they favor them.

This research made use of numerous HST observations and was partially 
supported by grants 10844 and 11291 from the Space Telescope Science 
Institute (STScI), which is operated by the Association of Universities 
for Research in Astronomy, Inc., under NASA contract NAS 5-26555. 
The data archive for the HST Treasury Program on Eta Carinae (HST program 
GO 9973) is available online at http://etacar.umn.edu/, partially 
supported by the University of Minnesota and STScI.  Our analyses 
used software tools developed for the Treasury Program by K.\ Davidson, 
K.\ Ishibashi, and J.~C.\ Martin. Calculations were performed with 
version 08.00 of the photoionization program Cloudy 
described by \citet{1998PASP...110..761F}.
Finally, we wish to thank the referee, N.\ Soker, for pointing out  
some logical gaps and pertinent references that we neglected in the 
original version. 
  

\newpage

\bibliographystyle{apj}
\bibliography{ms} 

\begin{thebibliography}{75}
\expandafter\ifx\csname natexlab\endcsname\relax\def\natexlab#1{#1}\fi

\bibitem[{{Aller} \& {Dunham}(1966)}]{1966ApJ...146..126A}
{Aller}, L.~H. \& {Dunham}, T.~J. 1966, \apj, 146, 126

\bibitem[{{Castelli} \& {Kurucz}(2004)}]{2004astro.ph..5087C}
{Castelli}, F. \& {Kurucz}, R.~L. 2004, ArXiv Astrophysics e-prints

\bibitem[{{Corcoran}(2005)}]{2005AJ....129.2018C}
{Corcoran}, M.~F. 2005, \aj, 129, 2018

\bibitem[{{Damineli}(1996)}]{1996ApJ...460L..49D}
{Damineli}, A. 1996, \apjl, 460, L49

\bibitem[{{Damineli} {et~al.}(2008{\natexlab{a}}){Damineli}, {Hillier},
  {Corcoran}, {Stahl}, {Groh}, {Arias}, {Teodoro}, {Morrell}, {Gamen},
  {Gonzalez}, {Leister}, {Levato}, {Levenhagen}, {Grosso}, {Colombo}, \&
  {Wallerstein}}]{2008MNRAS.386.2330D}
{Damineli}, A., {Hillier}, D.~J., {Corcoran}, M.~F., {Stahl}, O., {Groh},
  J.~H., {Arias}, J., {Teodoro}, M., {Morrell}, N., {Gamen}, R., {Gonzalez},
  F., {Leister}, N.~V., {Levato}, H., {Levenhagen}, R.~S., {Grosso}, M.,
  {Colombo}, J.~F.~A., \& {Wallerstein}, G. 2008{\natexlab{a}}, \mnras, 386,
  2330

\bibitem[{{Damineli} {et~al.}(2008{\natexlab{b}}){Damineli}, {Hillier},
  {Corcoran}, {Stahl}, {Levenhagen}, {Leister}, {Groh}, {Teodoro}, {Albacete
  Colombo}, {Gonzalez}, {Arias}, {Levato}, {Grosso}, {Morrell}, {Gamen},
  {Wallerstein}, \& {Niemela}}]{2008MNRAS.384.1649D}
{Damineli}, A., {Hillier}, D.~J., {Corcoran}, M.~F., {Stahl}, O., {Levenhagen},
  R.~S., {Leister}, N.~V., {Groh}, J.~H., {Teodoro}, M., {Albacete Colombo},
  J.~F., {Gonzalez}, F., {Arias}, J., {Levato}, H., {Grosso}, M., {Morrell},
  N., {Gamen}, R., {Wallerstein}, G., \& {Niemela}, V. 2008{\natexlab{b}},
  \mnras, 384, 1649

\bibitem[{{Damineli} {et~al.}(1998){Damineli}, {Stahl}, {Kaufer}, {Wolf},
  {Quast}, \& {Lopes}}]{1998A&AS..133..299D}
{Damineli}, A., {Stahl}, O., {Kaufer}, A., {Wolf}, B., {Quast}, G., \& {Lopes},
  D.~F. 1998, \aaps, 133, 299

\bibitem[{{Davidson}(1997)}]{1997NewA....2..387D}
{Davidson}, K. 1997, New Astronomy, 2, 387

\bibitem[{{Davidson}(1999)}]{1999ASPC..179..304D}
{Davidson}, K. 1999, in ASP Conf. Ser., Vol. 179, Eta Carinae at The
  Millennium, ed. J.~A. {Morse}, R.~M. {Humphreys}, \& A.~{Damineli}, 304

\bibitem[{{Davidson}(2006)}]{2006hstc.conf..247D}
{Davidson}, K. 2006, in The 2005 HST Calibration Workshop: Hubble After the
  Transition to Two-Gyro Mode, ed. A.~M. {Koekemoer}, P.~{Goudfrooij}, \& L.~L.
  {Dressel}, 247

\bibitem[{{Davidson} {et~al.}(1986){Davidson}, {Dufour}, {Walborn}, \&
  {Gull}}]{1986ApJ...305..867D}
{Davidson}, K., {Dufour}, R.~J., {Walborn}, N.~R., \& {Gull}, T.~R. 1986, \apj,
  305, 867

\bibitem[{{Davidson} {et~al.}(1997){Davidson}, {Ebbets}, {Johansson}, {Morse},
  \& {Hamann}}]{1997AJ....113..335D}
{Davidson}, K., {Ebbets}, D., {Johansson}, S., {Morse}, J.~A., \& {Hamann},
  F.~W. 1997, \aj, 113, 335

\bibitem[{{Davidson} {et~al.}(1995){Davidson}, {Ebbets}, {Weigelt},
  {Humphreys}, {Hajian}, {Walborn}, \& {Rosa}}]{1995AJ....109.1784D}
{Davidson}, K., {Ebbets}, D., {Weigelt}, G., {Humphreys}, R.~M., {Hajian},
  A.~R., {Walborn}, N.~R., \& {Rosa}, M. 1995, \aj, 109, 1784

\bibitem[{{Davidson} \& {Humphreys}(1997)}]{1997ARA&A..35....1D}
{Davidson}, K. \& {Humphreys}, R.~M. 1997, \araa, 35, 1

\bibitem[{{Davidson} {et~al.}(2005){Davidson}, {Martin}, {Humphreys},
  {Ishibashi}, {Gull}, {Stahl}, {Weis}, {Hillier}, {Damineli}, {Corcoran}, \&
  {Hamann}}]{2005AJ....129..900D}
{Davidson}, K., {Martin}, J., {Humphreys}, R.~M., {Ishibashi}, K., {Gull},
  T.~R., {Stahl}, O., {Weis}, K., {Hillier}, D.~J., {Damineli}, A., {Corcoran},
  M., \& {Hamann}, F. 2005, \aj, 129, 900

\bibitem[{{Davidson} \& {Netzer}(1979)}]{1979RvMP...51..715D}
{Davidson}, K. \& {Netzer}, H. 1979, Reviews of Modern Physics, 51, 715

\bibitem[{{Davidson} {et~al.}(2001){Davidson}, {Smith}, {Gull}, {Ishibashi}, \&
  {Hillier}}]{2001AJ....121.1569D}
{Davidson}, K., {Smith}, N., {Gull}, T.~R., {Ishibashi}, K., \& {Hillier},
  D.~J. 2001, \aj, 121, 1569

\bibitem[{{Dorland} {et~al.}(2004){Dorland}, {Currie}, \&
  {Hajian}}]{2004AJ....127.1052D}
{Dorland}, B.~N., {Currie}, D.~G., \& {Hajian}, A.~R. 2004, \aj, 127, 1052

\bibitem[{{Dufour} {et~al.}(1999){Dufour}, {Glover}, {Hester}, {Currie}, {van
  Orsow}, \& {Walter}}]{1999ASPC..179..134D}
{Dufour}, R.~J., {Glover}, T.~W., {Hester}, J.~J., {Currie}, D.~G., {van
  Orsow}, D., \& {Walter}, D.~K. 1999, in ASP Conf. Ser., Vol. 179, Eta Carinae
  at The Millennium, ed. J.~A. {Morse}, R.~M. {Humphreys}, \& A.~{Damineli},
  134

\bibitem[{{Ferland} {et~al.}(1998){Ferland}, {Korista}, {Verner}, {Ferguson},
  {Kingdon}, \& {Verner}}]{1998PASP...110..761F}
{Ferland}, G.~J., {Korista}, K.~T., {Verner}, D.~A., {Ferguson}, J.~W.,
  {Kingdon}, J.~B., \& {Verner}, E.~M. 1998, \pasp, 110, 761

\bibitem[{{Gaviola}(1953)}]{1953ApJ...118..234G}
{Gaviola}, E. 1953, \apj, 118, 234

\bibitem[{{Gull} {et~al.}(1999){Gull}, {Ishibashi}, {Davidson}, \& {The Cycle 7
  STIS Go Team}}]{1999ASPC..179..144G}
{Gull}, T.~R., {Ishibashi}, K., {Davidson}, K., \& {The Cycle 7 STIS Go Team}.
  1999, in ASP Conf. Ser., Vol. 179, Eta Carinae at The Millennium, ed. J.~A.
  {Morse}, R.~M. {Humphreys}, \& A.~{Damineli}, 144

\bibitem[{{Gull} {et~al.}(2001){Gull}, {Johannson}, \&
  {Davidson}}]{2001ASPC..242.....G}
{Gull}, T.~R., {Johannson}, S., \& {Davidson}, K., eds. 2001, ASP Conf. Ser.,
  Vol. 242, {Eta Carinae and Other Mysterious Stars: The Hidden Opportunities
  of Emission Spectroscopy.}

\bibitem[{{Gull} {et~al.}(2009){Gull}, {Nielsen}, {Corcoran}, {Madura},
  {Owocki}, {Russell}, {Hillier}, {Hamaguchi}, {Kober}, {Weis}, {Stahl}, \&
  {Okazaki}}]{2009MNRAS.396.1308G}
{Gull}, T.~R., {Nielsen}, K.~E., {Corcoran}, M.~F., {Madura}, T.~I., {Owocki},
  S.~P., {Russell}, C.~M.~P., {Hillier}, D.~J., {Hamaguchi}, K., {Kober},
  G.~V., {Weis}, K., {Stahl}, O., \& {Okazaki}, A.~T. 2009, \mnras, 396, 1308

\bibitem[{{Hamann} {et~al.}(1999){Hamann}, {Davidson}, {Ishibashi}, \&
  {Gull}}]{1999ASPC..179..116H}
{Hamann}, F., {Davidson}, K., {Ishibashi}, K., \& {Gull}, T.~R. 1999, in ASP
  Conf. Ser., Vol. 179, Eta Carinae at The Millennium, ed. J.~A. {Morse}, R.~M.
  {Humphreys}, \& A.~{Damineli}, 116

\bibitem[{{Hamann} {et~al.}(1994){Hamann}, {Depoy}, {Johansson}, \&
  {Elias}}]{1994ApJ...422..626H}
{Hamann}, F., {Depoy}, D.~L., {Johansson}, S., \& {Elias}, J. 1994, \apj, 422,
  626

\bibitem[{{Hillier} {et~al.}(2001){Hillier}, {Davidson}, {Ishibashi}, \&
  {Gull}}]{2001ApJ...553..837H}
{Hillier}, D.~J., {Davidson}, K., {Ishibashi}, K., \& {Gull}, T. 2001, \apj,
  553, 837

\bibitem[{{Hofmann} \& {Weigelt}(1988)}]{1988A&A...203L..21H}
{Hofmann}, K.-H. \& {Weigelt}, G. 1988, \aap, 203, L21

\bibitem[{{Humphreys} \& {Stanek}(2005)}]{2005ASPC..332.....H}
{Humphreys}, R. \& {Stanek}, K., eds. 2005, ASP Conf. Ser., Vol. 332, {The Fate
  of the Most Massive Stars}

\bibitem[{{Humphreys} {et~al.}(2008){Humphreys}, {Davidson}, \&
  {Koppelman}}]{2008AJ....135.1249H}
{Humphreys}, R.~M., {Davidson}, K., \& {Koppelman}, M. 2008, \aj, 135, 1249

\bibitem[{{Iben}(1999)}]{1999ASPC..179..367I}
{Iben}, Jr., I. 1999, in ASP Conf. Ser., Vol. 179, Eta Carinae at The
  Millennium, ed. J.~A. {Morse}, R.~M. {Humphreys}, \& A.~{Damineli}, 367

\bibitem[{{Ishibashi}(2001)}]{2001ASPC..242...53I}
{Ishibashi}, K. 2001, in ASP Conf. Ser., Vol. 242, Eta Carinae and Other
  Mysterious Stars: The Hidden Opportunities of Emission Spectroscopy, ed.
  T.~R. {Gull}, S.~{Johannson}, \& K.~{Davidson}, 53

\bibitem[{{Ishibashi} {et~al.}(1999{\natexlab{a}}){Ishibashi}, {Corcoran},
  {Davidson}, {Swank}, {Petre}, {Drake}, {Damineli}, \&
  {White}}]{1999ApJ...524..983I}
{Ishibashi}, K., {Corcoran}, M.~F., {Davidson}, K., {Swank}, J.~H., {Petre},
  R., {Drake}, S.~A., {Damineli}, A., \& {White}, S. 1999{\natexlab{a}}, \apj,
  524, 983

\bibitem[{{Ishibashi} {et~al.}(1999{\natexlab{b}}){Ishibashi}, {Davidson},
  {Corcoran}, {Drake}, {Swank}, \& {Petre}}]{1999ASPC..179..266I}
{Ishibashi}, K., {Davidson}, M.~F., {Corcoran}, K., {Drake}, S.~A., {Swank},
  J.~H., \& {Petre}, R. 1999{\natexlab{b}}, in ASP Conf. Ser., Vol. 179, Eta
  Carinae at The Millennium, ed. J.~A. {Morse}, R.~M. {Humphreys}, \&
  A.~{Damineli}, 266

\bibitem[{{Ishibashi} {et~al.}(2003){Ishibashi}, {Gull}, {Davidson}, {Smith},
  {Lanz}, {Lindler}, {Feggans}, {Verner}, {Woodgate}, {Kimble}, {Bowers},
  {Kraemer}, {Heap}, {Danks}, {Maran}, {Joseph}, {Kaiser}, {Linsky}, {Roesler},
  \& {Weistrop}}]{2003AJ....125.3222I}
{Ishibashi}, K., {Gull}, T.~R., {Davidson}, K., {Smith}, N., {Lanz}, T.,
  {Lindler}, D., {Feggans}, K., {Verner}, E., {Woodgate}, B.~E., {Kimble},
  R.~A., {Bowers}, C.~W., {Kraemer}, S., {Heap}, S.~R., {Danks}, A.~C.,
  {Maran}, S.~P., {Joseph}, C.~L., {Kaiser}, M.~E., {Linsky}, J.~L., {Roesler},
  F., \& {Weistrop}, D. 2003, \aj, 125, 3222

\bibitem[{{Johansson} {et~al.}(2006){Johansson}, {Hartman}, \&
  {Letokhov}}]{2006A&A...452..253J}
{Johansson}, S., {Hartman}, H., \& {Letokhov}, V.~S. 2006, \aap, 452, 253

\bibitem[{{Johansson} {et~al.}(2000){Johansson}, {Zethson}, {Hartman},
  {Ekberg}, {Ishibashi}, {Davidson}, \& {Gull}}]{2000A&A...361..977J}
{Johansson}, S., {Zethson}, T., {Hartman}, H., {Ekberg}, J.~O., {Ishibashi},
  K., {Davidson}, K., \& {Gull}, T. 2000, \aap, 361, 977

\bibitem[{{Kashi} \& {Soker}(2008)}]{2008MNRAS.390.1751K}
{Kashi}, A. \& {Soker}, N. 2008, \mnras, 390, 1751

\bibitem[{{Kashi} \& {Soker}(2009{\natexlab{a}})}]{2009NewA...14...11K}
{Kashi}, A. \& {Soker}, N. 2009{\natexlab{a}}, New Astronomy, 14, 11

\bibitem[{{Kashi} \& {Soker}(2009{\natexlab{b}})}]{2009MNRAS.394..923K}
{Kashi}, A. \& {Soker}, N. 2009{\natexlab{b}}, \mnras, 394, 923

\bibitem[{{Kashi} \& {Soker}(2009{\natexlab{c}})}]{2009MNRAS.397.1426K}
{Kashi}, A. \& {Soker}, N. 2009{\natexlab{c}}, \mnras, 397, 1426

\bibitem[{{Kudritzki} \& {Puls}(2000)}]{2000ARA&A..38..613K}
{Kudritzki}, R.-P. \& {Puls}, J. 2000, \araa, 38, 613

\bibitem[{{Lanz} \& {Hubeny}(2003)}]{2003ApJS..146..417L}
{Lanz}, T. \& {Hubeny}, I. 2003, \apjs, 146, 417

\bibitem[{{Martin} {et~al.}(2006{\natexlab{a}}){Martin}, {Davidson}, {Hamann},
  {Stahl}, \& {Weis}}]{2006PASP..118..697M}
{Martin}, J.~C., {Davidson}, K., {Hamann}, F., {Stahl}, O., \& {Weis}, K.
  2006{\natexlab{a}}, \pasp, 118, 697

\bibitem[{{Martin} {et~al.}(2006{\natexlab{b}}){Martin}, {Davidson},
  {Humphreys}, {Hillier}, \& {Ishibashi}}]{2006ApJ...640..474M}
{Martin}, J.~C., {Davidson}, K., {Humphreys}, R.~M., {Hillier}, D.~J., \&
  {Ishibashi}, K. 2006{\natexlab{b}}, \apj, 640, 474

\bibitem[{{Martin} {et~al.}(2009){Martin}, {Davidson}, {Humphreys}, \&
  {Mehner}}]{2009arXiv0908.1627M}
{Martin}, J.~C., {Davidson}, K., {Humphreys}, R.~M., \& {Mehner}, A. 2009,
  ArXiv e-prints

\bibitem[{{Martin} {et~al.}(2006{\natexlab{c}}){Martin}, {Davidson}, \&
  {Koppelman}}]{2006AJ....132.2717M}
{Martin}, J.~C., {Davidson}, K., \& {Koppelman}, M.~D. 2006{\natexlab{c}}, \aj,
  132, 2717

\bibitem[{{Martins} {et~al.}(2005){Martins}, {Schaerer}, \&
  {Hillier}}]{2005A&A...436.1049M}
{Martins}, F., {Schaerer}, D., \& {Hillier}, D.~J. 2005, \aap, 436, 1049

\bibitem[{{Morse} {et~al.}(1999){Morse}, {Humphreys}, \&
  {Damineli}}]{1999ASPC..179.....M}
{Morse}, J.~A., {Humphreys}, R.~M., \& {Damineli}, A., eds. 1999, ASP Conf.
  Ser., Vol. 179, {Eta Carinae At The Millennium}

\bibitem[{{Nahar} \& {Pradhan}(1996)}]{1996A&AS..119..509N}
{Nahar}, S.~N. \& {Pradhan}, A.~K. 1996, \aaps, 119, 509

\bibitem[{{Okazaki} {et~al.}(2008){Okazaki}, {Owocki}, {Russell}, \&
  {Corcoran}}]{2008MNRAS.388L..39O}
{Okazaki}, A.~T., {Owocki}, S.~P., {Russell}, C.~M.~P., \& {Corcoran}, M.~F.
  2008, \mnras, 388, L39

\bibitem[{{Osterbrock} \& {Ferland}(2006)}]{2006agna.book.....O}
{Osterbrock}, D.~E. \& {Ferland}, G.~J. 2006, {Astrophysics of gaseous nebulae
  and active galactic nuclei}, ed. D.~E. {Osterbrock} \& G.~J. {Ferland}

\bibitem[{{Parkin} {et~al.}(2009){Parkin}, {Pittard}, {Corcoran}, {Hamaguchi},
  \& {Stevens}}]{2009MNRAS.394.1758P}
{Parkin}, E.~R., {Pittard}, J.~M., {Corcoran}, M.~F., {Hamaguchi}, K., \&
  {Stevens}, I.~R. 2009, \mnras, 394, 1758

\bibitem[{{Pauldrach} {et~al.}(2001){Pauldrach}, {Hoffmann}, \&
  {Lennon}}]{2001A&A...375..161P}
{Pauldrach}, A.~W.~A., {Hoffmann}, T.~L., \& {Lennon}, M. 2001, \aap, 375, 161

\bibitem[{{Pittard} \& {Corcoran}(2002)}]{2002A&A...383..636P}
{Pittard}, J.~M. \& {Corcoran}, M.~F. 2002, \aap, 383, 636

\bibitem[{{Quinet}(1996)}]{1996A&AS..116..573Q}
{Quinet}, P. 1996, \aaps, 116, 573

\bibitem[{{Schaerer} \& {de Koter}(1997)}]{1997A&A...322..598S}
{Schaerer}, D. \& {de Koter}, A. 1997, \aap, 322, 598

\bibitem[{{Smith} {et~al.}(2002){Smith}, {Norris}, \&
  {Crowther}}]{2002MNRAS.337.1309S}
{Smith}, L.~J., {Norris}, R.~P.~F., \& {Crowther}, P.~A. 2002, \mnras, 337,
  1309

\bibitem[{{Smith}(2006)}]{2006ApJ...644.1151S}
{Smith}, N. 2006, \apj, 644, 1151

\bibitem[{{Smith} {et~al.}(2003){Smith}, {Davidson}, {Gull}, {Ishibashi}, \&
  {Hillier}}]{2003ApJ...586..432S}
{Smith}, N., {Davidson}, K., {Gull}, T.~R., {Ishibashi}, K., \& {Hillier},
  D.~J. 2003, \apj, 586, 432

\bibitem[{{Smith} {et~al.}(2000){Smith}, {Morse}, {Davidson}, \&
  {Humphreys}}]{2000AJ....120..920S}
{Smith}, N., {Morse}, J.~A., {Davidson}, K., \& {Humphreys}, R.~M. 2000, \aj,
  120, 920

\bibitem[{{Smith} {et~al.}(2004){Smith}, {Morse}, {Gull}, {Hillier}, {Gehrz},
  {Walborn}, {Bautista}, {Collins}, {Corcoran}, {Damineli}, {Hamann},
  {Hartman}, {Johansson}, {Stahl}, \& {Weis}}]{2004ApJ...605..405S}
{Smith}, N., {Morse}, J.~A., {Gull}, T.~R., {Hillier}, D.~J., {Gehrz}, R.~D.,
  {Walborn}, N.~R., {Bautista}, M., {Collins}, N.~R., {Corcoran}, M.~F.,
  {Damineli}, A., {Hamann}, F., {Hartman}, H., {Johansson}, S., {Stahl}, O., \&
  {Weis}, K. 2004, \apj, 605, 405

\bibitem[{{Soker}(2007)}]{2007ApJ...661..490S}
{Soker}, N. 2007, \apj, 661, 490

\bibitem[{{Thackeray}(1953)}]{1953MNRAS.113..211T}
{Thackeray}, A.~D. 1953, \mnras, 113, 211

\bibitem[{{Thackeray}(1967)}]{1967MNRAS.135...51T}
{Thackeray}, A.~D. 1967, \mnras, 135, 51

\bibitem[{{Verner} {et~al.}(2005){Verner}, {Bruhweiler}, \&
  {Gull}}]{2005ApJ...624..973V}
{Verner}, E., {Bruhweiler}, F., \& {Gull}, T. 2005, \apj, 624, 973

\bibitem[{{Viotti} {et~al.}(1989){Viotti}, {Rossi}, {Cassatella}, {Altamore},
  \& {Baratta}}]{1989ApJS...71..983V}
{Viotti}, R., {Rossi}, L., {Cassatella}, A., {Altamore}, A., \& {Baratta},
  G.~B. 1989, \apjs, 71, 983

\bibitem[{{Walborn}(1995)}]{1995RMxAC...2...51W}
{Walborn}, N.~R. 1995, in \rmxaa Conference Series, Vol.~2, \rmxaa Conference
  Series, ed. V.~{Niemela}, N.~{Morrell}, \& A.~{Feinstein}, 51

\bibitem[{{Weigelt} {et~al.}(1995){Weigelt}, {Albrecht}, {Barbieri}, {Blades},
  {Boksenberg}, {Crane}, {Davidson}, {Deharveng}, {Disney}, {Jakobsen},
  {Kamperman}, {King}, {Macchetto}, {Mackay}, {Paresce}, {Baxter},
  {Greenfield}, {Jedrzejewski}, {Nota}, \& {Sparks}}]{1995RMxAC...2...11W}
{Weigelt}, G., {Albrecht}, R., {Barbieri}, C., {Blades}, J.~C., {Boksenberg},
  A., {Crane}, P., {Davidson}, K., {Deharveng}, J.~M., {Disney}, M.~J.,
  {Jakobsen}, P., {Kamperman}, T.~M., {King}, I.~R., {Macchetto}, F., {Mackay},
  C.~D., {Paresce}, F., {Baxter}, D., {Greenfield}, P., {Jedrzejewski}, R.,
  {Nota}, A., \& {Sparks}, W.~B. 1995, in \rmxaa Conference Series, Vol.~2,
  \rmxaa Conference Series, ed. V.~{Niemela}, N.~{Morrell}, \& A.~{Feinstein},
  11

\bibitem[{{Weigelt} \& {Ebersberger}(1986)}]{1986A&A...163L...5W}
{Weigelt}, G. \& {Ebersberger}, J. 1986, \aap, 163, L5

\bibitem[{{Whitelock} {et~al.}(1983){Whitelock}, {Feast}, {Carter}, {Roberts},
  \& {Glass}}]{1983MNRAS.203..385W}
{Whitelock}, P.~A., {Feast}, M.~W., {Carter}, B.~S., {Roberts}, G., \& {Glass},
  I.~S. 1983, \mnras, 203, 385

\bibitem[{{Whitelock} {et~al.}(1994){Whitelock}, {Feast}, {Koen}, {Roberts}, \&
  {Carter}}]{1994MNRAS.270..364W}
{Whitelock}, P.~A., {Feast}, M.~W., {Koen}, C., {Roberts}, G., \& {Carter},
  B.~S. 1994, \mnras, 270, 364

\bibitem[{{Zanella} {et~al.}(1984){Zanella}, {Wolf}, \&
  {Stahl}}]{1984A&A...137...79Z}
{Zanella}, R., {Wolf}, B., \& {Stahl}, O. 1984, \aap, 137, 79

\bibitem[{{Zethson}(2001)}]{2001PhDT.........1Z}
{Zethson}, T. 2001, PhD thesis, AA(LUNDS UNIVERSITET (SWEDEN))

\bibitem[{{Zethson} {et~al.}(1999){Zethson}, {Johansson}, {Davidson},
  {Humphreys}, {Ishibashi}, \& {Ebbets}}]{1999A&A...344..211Z}
{Zethson}, T., {Johansson}, S., {Davidson}, K., {Humphreys}, R.~M.,
  {Ishibashi}, K., \& {Ebbets}, D. 1999, \aap, 344, 211

\end{thebibliography}

\newpage 


\begin{deluxetable}{lcrc}    
\tabletypesize{\scriptsize} 
\tablecaption{Times of STIS observations used in the emission maps
   \label{tab:table1}}   
\tablewidth{0pt} \tablehead{ \colhead{Date (UT)} & \colhead{MJD}
     &  \colhead{Slit PA\tablenotemark{a}}  
     &  \colhead{$N_{\mathrm{slit}}$\tablenotemark{b}}  }   
\startdata
   1998 Nov 25\tablenotemark{c}  
                          &  51142  &  227{\degree}  &  1  \\    
   1999 Feb 22  &  51231  &  332{\degree}  &  1  \\      
   2000 Mar 21  &  51624  &  332{\degree}  &  1  \\ 
   2001 Apr 18  &  52017  &   22{\degree}  &  1  \\  
   2001 Oct 01\tablenotemark{c}
                          &  52183  &  165{\degree}  &  1  \\     
   2002 Jan 20  &  52294  &  278{\degree}  &  1  \\  
   2002 Jul 05  &  52460  &   69{\degree}  &  2  \\   
   2003 Feb 13  &  52683  &  303{\degree}  &  1  \\   
   2003 Mar 29  &  52727  &  332{\degree}  &  1  \\  
   2003 May 05  &  52764  &   27{\degree}  &  1  \\  
   2003 May 17  &  52776  &   38{\degree}  &  2  \\  
   2003 Jun 02  &  52792  &   62{\degree}  &  2  \\   
\enddata 
\tablenotetext{a}{PA $\pm$ 180\degree \ gives same spatial coverage.}  
\tablenotetext{b}{The star was observed on every occasion, but 
       sometimes there was also a second slit position offset about 
       0.25{\arcsec} NW of the star.}  
\tablenotetext{c}{There were no [Fe~III] obs.\ on this date.}     
\end{deluxetable}  


\newpage

\begin{deluxetable}{ccc}     
\tabletypesize{\scriptsize} 
\tablecaption{Chemical compositions\label{tab:table2}}   
\tablewidth{0pt} 
\tablehead{ & 
   \multicolumn{2}{c}{\phantom{~~}$\mathrm{log_{10} ~N(element)/N(H)}$ \phantom{~} }  \\ 
     & \colhead{\phantom{~~}Comp.\ A\tablenotemark{a}\phantom{~}} &  \colhead{\phantom{~~}Comp.\ B\tablenotemark{b}\phantom{~}}}   
\startdata
    He    &    $-0.70$    &    $-0.70$   \\
    C     &    $-5.00$    &    $-4.70$   \\  
    N     &    $-3.10$    &    $-2.80$   \\  
    O     &    $-5.00$    &    $-4.70$   \\  
    Ne    &    $-4.00$    &    $-3.70$   \\  
    Si    &    $-4.46$    &    $-4.16$   \\  
    S     &    $-4.74$    &    $-4.44$   \\  
    Ar    &    $-5.60$    &    $-5.30$   \\  
    Fe    &    $-4.55$    &    $-4.25$   \\  
\enddata 
\tablenotetext{a}{Default solar composition as used in Cloudy, except He, C, N, O.}  
\tablenotetext{b}{CNO fraction somewhat larger than solar.} 
\end{deluxetable}  



\newpage


\begin{deluxetable}{lccccc}   
\tabletypesize{\scriptsize} 
\tablecaption{Selected high-excitation emission lines in Weigelt blob C
    \label{tab:table3}} 
\tablewidth{0pt} \tablehead{
\colhead{Spectrum} &
\colhead{$\mathrm{\lambda_{vac}}$} &
\colhead{I.P.'s\tablenotemark{a}} &
\colhead{EW} &
\colhead{$I_\mathrm{obs}$\tablenotemark{b} }  & 
\colhead{$I_\mathrm{corr}$\tablenotemark{c} } \\ 
\colhead{}&
\colhead{(\AA)} &
\colhead{(eV)} &
\colhead{(\AA)}&
\colhead{} &
\colhead{}}
\startdata
{[}Ne III{]} & 3869.85 & 40.96--63.45   & 10.24  &  3.63  &  1.00  \\  
Si III{]}    & 1892.03 & 16.34--33.49   &  3.84  &  2.78  &  2.10  \\  
{[}Ar III{]} & 3110.08 & 27.63--40.74   &  0.42  &  0.17  &  0.06  \\
{[}Ar III{]} & 5193.26 & 27.63--40.74   &  0.39  &  0.11  &  0.024  \\
{[}Ar III{]} & 7137.76 & 27.63--40.74   &  6.36  &  2.24  &  0.40  \\
{[}Ar III{]} & 7753.24 & 27.63--40.74   &  1.93  &  0.59  &  0.10  \\
He I & 4027.33 & 24.59--54.42\tablenotemark{d} &  2.42 & 0.69 & 0.18 \\
He I & 6680.00 & 24.59--54.42\tablenotemark{d} &  5.44 & 2.54 & 0.47 \\
He I & 7067.20 & 24.59--54.42\tablenotemark{d} & 16.19 & 10.1 & 1.80 \\
{[}S III{]}  & 6313.81 & 23.33--34.83   &  7.72  &  2.39  &  0.45  \\
{[}S III{]}  & 9071.11 & 23.33--34.83   &  4.06  &  1.46  &  0.23  \\
{[}S III{]}  & 9533.23 & 23.33--34.83   &  5.49  &  2.49  &  0.39  \\
{[}Fe III{]} & 4659.35 & 16.18--30.65   &  4.27  &  1.97  &  0.46  \\
{[}Fe III{]} & 4702.85 & 16.18--30.65   &  3.21  &  0.88  &   0.21    
\enddata
\tablerefs{Atomic data from http://physics.nist.gov/ PhysRefData/ASD/.}  
\tablenotetext{a} {Relevant ionization potentials.   For instance,  
   Ne$^+$ and Ne$^{++}$ have ionization potentials 40.96 and 63.45 eV.}
\tablenotetext{b} {In units of $10^{-12}$ erg cm$^{-2}$ s$^{-1}$ within the 
   0.1{\arcsec} sampled area.}
\tablenotetext{c} {De-reddened intrinsic strength relative to 
   {[}Ne~III{]} ${\lambda}3870$.}
\tablenotetext{d}{Recombination spectrum created in He$^+$ zone.}
\end{deluxetable}


\newpage

\begin{deluxetable}{lccccc}   
\tabletypesize{\scriptsize} 
\tablecaption{Comparison between different stellar atmosphere models\tablenotemark{a}\label{tab:table4}} 
\tablewidth{0pt} \tablehead{
\colhead{Model} &
\colhead{$L$} &
\colhead{log $L/L_{\odot}$} &
\colhead{log $Q_{\mathrm{H}}$} &
\colhead{log $Q_{\mathrm{He}}$} &
\colhead{log $Q_{\mathrm{Ne III}}$} \\ 
\colhead{}&
\colhead{$\mathrm{(erg/s^{-1})}$}&
\colhead{} &
\colhead{$\mathrm{(photons~s^{-1})}$}&
\colhead{$\mathrm{(photons~s^{-1})}$}&
\colhead{$\mathrm{(photons~s^{-1})}$}
}
\startdata
Atlas  & 3.67e+38  & 4.98 &  48.72 & 48.27 & 46.09 \\  
  \\  
CoStar  & 4.25e+38 & 5.04 &   48.90 & 48.70  &  48.32 \\ 
 \\ 
Tlusty &  3.27e+38 & 4.93 &  48.76 & 48.40 & 46.33 \\
\\
WM-basic & 3.99e+38  &  5.02 & 48.82 & 48.51 & 47.04  
\enddata
\tablenotetext{a}{$T_\mathrm{eff}$ = 40000 K, $n_\mathrm{H}$= 10$^{6}$ cm$^{-3}$ and log $U_\mathrm{H} = -1$}
\end{deluxetable}


\begin{deluxetable}{lcccccccc}  
\tabletypesize{\scriptsize} 
\tablecaption{WM-basic model predictions of $T_\mathrm{eff}$ for secondary star and composition A\label{tab:table5}} 
\tablewidth{0pt} \tablehead{
\colhead{Density} &
\colhead{log $U_\mathrm{H}$} &
\colhead{log $Q_\mathrm{H}$} &
\colhead{$T_\mathrm{eff}$(Ar 7138)\tablenotemark{a}} &
\colhead{$T_\mathrm{eff}$(He 6680)\tablenotemark{a}} &
\colhead{$T_\mathrm{eff}$(Si 1892)\tablenotemark{a}} &
\colhead{$T_\mathrm{avg}$} &
\colhead{$\Delta T_\mathrm{avg}$}&
\colhead{log $L/L_\mathrm{\odot}$} 
\\ 
\colhead{($\mathrm{cm^{-3}}$)}&
\colhead{}&
\colhead{$\mathrm{(photons~s^{-1})}$}&
\colhead{(K)} &
\colhead{(K)}&
\colhead{(K)}&
\colhead{(K)}&
\colhead{(K)}&
\colhead{}
}
\startdata
$\mathrm{10^{5}}$  & -2.14 & 46.44 & 48700 & 40300 & 37600 & 42200\tablenotemark{c} & 5800 & 2.84\\ 
$\mathrm{10^{5}}$  & -2 & 46.58 &48700 &  38900& 37200 & 41600\tablenotemark{b,c} & 6200 & 2.99\\  
$\mathrm{10^{5}}$  & -1& 47.58 &42300 &  35900 & 35100 & 37800\tablenotemark{b,c} & 3900 & 4.08\\  
$\mathrm{10^{5}}$  & 0  & 48.58 & 39100 &  34200 & 33900 & 35700\tablenotemark{b} & 2900 & 5.16 \\  
$\mathrm{10^{5}}$  & 1 & 49.58  & 37700 &  33300 & 33500 & 34800\tablenotemark{b} & 2500 &6.22  \\  
$\mathrm{10^{5}}$  & 2 & 50.58 & 36800 &  33000 & 33100 & 34300\tablenotemark{b,c} & 2200 & 7.31\\ 
 \\  
$\mathrm{10^{6}}$  & -2 & 47.58  & 43400 &   37600 & 41100  &  40700\tablenotemark{c} & 2900 & 4.00\\ 
$\mathrm{10^{6}}$  & -1 & 48.58 & 38100 &   33900 & 36700  &  36500 & 2300 & 5.13 \\ 
$\mathrm{10^{6}}$  & -0.67 & 48.91 & 37200 &   33600 & 37200  &  36000 & 2100 &  5.48\\ 
$\mathrm{10^{6}}$  & 0 & 49.58 & 36600 &   32800 & 36600  &  35300\tablenotemark{b} & 2200 & 6.18  \\ 
$\mathrm{10^{6}}$  & 1 &50.58  & 36200 &   32600 & 36300  &  35000\tablenotemark{b,c} & 2100 & 7.19\\ 
$\mathrm{10^{6}}$  & 2  & 51.58 & 36600 &   32800 & 35900  &  35100\tablenotemark{b,c} & 2000 & 8.19\\ 
 \\ 
$\mathrm{10^{7}}$ &  -2 & 48.58 & 39400 &  38000 & 51900 & 43100 & 7700 & 4.97 \\
$\mathrm{10^{7}}$ &  -1 & 49.58 & 36100 &  34900 & 43700 & 38200 & 4800 & 6.06 \\
$\mathrm{10^{7}}$ &  0 & 50.58 & 36000 &  34500 & 38700 & 36400\tablenotemark{c} & 2100 & 7.13\\
$\mathrm{10^{7}}$ &  0.24 & 50.82 & 35700 &  34500 & 38400 & 36200\tablenotemark{c} & 2000 & 7.38\\
$\mathrm{10^{7}}$ &   1 &  51.58  &  35500 &  34900 &  38100 &  36200\tablenotemark{b,c} &  1700 &  8.14\\
$\mathrm{10^{7\ast}}$ &  2 & 52.58 & 35600 &  34900 & 37800 & 36100\tablenotemark{b,c} & 1500 & 9.14
\enddata
\tablenotetext{a}{Relative to $[$Ne III$]$ \lm3870 \AA.}
\tablenotetext{b}{Models with He II region larger than $\mathrm{4\times 10^{15}~cm}$}
\tablenotetext{c}{Models with log $L/L_\mathrm{\odot}$ smaller then 5 or larger than 6.}
\end{deluxetable}


\end{document}